\documentclass[pra,amsmath,amssymb]{revtex4}

\usepackage{natbib}

\usepackage{graphics}
\usepackage{rotating}
\usepackage{epsfig}
\usepackage{color}

\def\inbar{\,\vrule height1.5ex width.4pt depth0pt}
\def\IR{\relax{\rm I\kern-.18em R}}
\def\IC{\relax\hbox{$\inbar\kern-.3em{X\rm C}$}}

\begin{document}
\title{Ultracold Feshbach Molecules}

\author{Francesca Ferlaino}
\author{Steven Knoop}
\author{Rudolf Grimm}

\affiliation{Institute of Experimental Physics and Center for
Quantum Physics, University of Innsbruck, 6020 Innsbruck, Austria,} \affiliation{Institute for Quantum Optics
and Quantum Information, Austrian Academy of Sciences,
6020 Innsbruck, Austria}

\maketitle

\tableofcontents

\hyphenation{Fesh-bach}

\section{Introduction}
\label{sec:intro}

The {\em coldest molecules} available for experiments are diatomic molecules produced by association techniques in ultracold atomic gases. The basic idea is to bind atoms together when they collide at extremely low kinetic energies. If any release of internal energy is avoided in this process, the molecular gas just inherits the ultralow temperature of the atomic gas, which can in practice be as low as a few nanokelvin. If moreover a {\em single molecular quantum state} is selectively populated, there will be no increase of the system's entropy. A very efficient method to implement such a scenario in quantum-degenerate gases, Bose-Einstein condensates and Fermi gases, relies on the magnetically controlled association via Feshbach resonances. Such resonances serve as the entrance gate into the molecular world and allow for the conversion of atomic into molecular quantum gases.

In this Chapter, we give an introduction into experiments with {\em Feshbach molecules} and their various applications. Our illustrative examples are mainly based on work performed at Innsbruck University, but the reader will also find references to related work of other laboratories.

\subsection{Ultracold atoms and quantum gases}

The development of techniques for cooling and trapping of atoms has lead to great advances in physics, which have already been recognized by two Nobel prizes. In 1997 the prize was jointly awarded to Steven Chu, Claude Cohen-Tannoudji and William D. Phillips ``for their developments of methods to cool and trap atoms with laser light'' \cite{Chu1998nlt, Cohentannoudji1998nlm, Phillips1998nll}. In 2001, Eric A. Cornell, Wolfgang Ketterle and Carl E. Wieman jointly received the Nobel prize ``for the achievement of Bose-Einstein condensation in dilute gases of alkali atoms, and for early fundamental studies of the properties of the condensates'' \cite{Cornell2002nlb, Ketterle2002nlw}.

Laser cooling techniques allow researchers to prepare atoms in the microkelvin range. The key techniques have been developed since the early 80's and a detailed account can be found in Refs.\ \cite{Varenna1991, Metcalf1999book}. In brief, resonance radiation pressure is used to decelerate atoms from a thermal beam to velocities low enough to be captured into a so-called magneto-optical trap or, alternatively, low-velocity atoms can be captured into such a trap directly from a vapor. In the trap the atoms are further cooled by Doppler cooling to temperatures of the order of one millikelvin. Then, for some species, sub-Doppler cooling methods can applied reducing the temperatures deep into the microkelvin range. Typical laser-cooled clouds contain up to about 10$^{10}$ atoms, and the atomic number densities are of the order of $10^{11}$\,cm$^{-3}$.
Such laser-cooled atoms have found numerous applications; a particularly important one is the realization of ultraprecise atomic clocks \cite{Bize2005cac, Hollberg2005ofw}.

The quantum-degenerate regime requires atomic de Broglie wavelengths to be similar to or larger than the interparticle spacing. As in laser-cooled clouds the phase-space densities are several orders of magnitude too low to reach this condition, further cooling needs to be applied. The method of choice is {\em evaporative cooling} \cite{Ketterle1997eco}, which can be applied in conservative traps like magnetic traps. 
Evaporative cooling relies on the selective removal of the most energetic atoms in combination with thermal equilibration of the sample by elastic collisions. The process can be forced by continuously lowering the trap depth. By trading one order of magnitude in the particle number one can gain up to typically three orders of magnitude in phase-space densities, and eventually reach the quantum-degenerate regime. The typical conditions are then a trapped gas of roughly a million atoms with densities of the order of 10$^{14}$\,cm$^{-3}$ and temperatures in the nanokelvin range.

Quantum degeneracy has been achieved with bosonic and fermionic atoms. The attainment of Bose-Einstein condensation (BEC) in dilute ultracold gases marked the starting point of a new era in physics \cite{Anderson1995oob, Bradley1995eob, Davis1995bec}, and degenerate Fermi gases entered the stage a few years later \cite{DeMarco1999oof, Schreck2001qbe, Truscott2001oof}. The reader may be referred to the proceedings of the Varenna summer schools in 1998 and 2006 \cite{Varenna1998, Varenna2006}, which describe these exciting developments. For reviews on the theory of degenerate quantum gases of bosons and fermions see \cite{Dalfovo1999tob} and \cite{Giorgini2007tou}, respectively, and the textbooks \cite{stringaribook} and \cite{pethickbook}.
The trapping environment plays an important role for the applications of ultracold matter, including the present one to create molecules. Most of the experiments on molecule formation are performed with {\em optical dipole traps} \cite{Grimm2000odt} for two main reasons. Such optical traps can confine atoms in any Zeeman or hyperfine substate of the electronic ground state; this includes the particularly interesting lowest internal state, which cannot be trapped magnetically. Moreover, optical dipole traps leave full freedom to apply any external magnetic fields, whereas in magnetic traps the application of additional magnetic fields would substantially affect the trapping geometry. A special optical trapping environment can be generated in {\em optical lattices} \cite{Bloch2005uqg, Greiner2008cmp}. Such lattices are created in optical standing-wave patterns, and in the three-dimensional case they provide an array of wavelength-sized microtraps, each of which may contain a single molecule.

The new field of ultracold molecular quantum gases rapidly emerged in 2002/03, when several groups reported on the formation of Feshbach molecules in BECs of $^{85}$Rb \cite{Donley2002amc}, $^{133}$Cs \cite{Herbig2003poa}, $^{87}$Rb \cite{Durr2004oom}, $^{23}$Na \cite{Xu2003foq}, as well as in degenerate or near-degenerate Fermi gases of $^{40}$K \cite{Regal2003cum} and $^6$Li \cite{Strecker2003coa, Cubizolles2003pol, Jochim2003pgo}. Signatures of molecules in a BEC had been seen before using a photoassociative technique \cite{Wynar2000mia}. In the fall of 2003, the fast progress culminated in the creation of molecular Bose-Einstein condensates (mBEC) in atomic Fermi gases \cite{Jochim2003bec, Greiner2003eoa, Zwierlein2003oob}. Heteronuclear Feshbach molecules entered the stage a few years later.
So far, such molecules have been produced in Bose-Fermi mixtures of $^{87}$Rb-$^{40}$K \cite{Ospelkaus2006uhm}, and in Bose-Bose mixtures of $^{85}$K-$^{87}$Rb \cite{Papp2006ooh} and $^{41}$K-$^{87}$Rb \cite{Weber2008aou}. The field of ultracold Feshbach molecules is developing rapidly, and there is great interest in homo- and heteronuclear molecules both in weakly bound and deeply bound regimes.

\subsection{Basic physics of a Feshbach resonance}
\label{ssec:feshres}

Feshbach resonances represent the essential tool to control the interaction between the atoms in ultracold gases, which has been the key to many breakthroughs. Here we briefly outline the basic physics of a Feshbach resonance and its connection to the underlying near-threshold molecular structure. For more detailed discussions of the theoretical background see Chapters 6 and 11.
The reader is also referred to two recent review articles \cite{Kohler2006poc, Chin2008fri}.

\begin{figure}
\includegraphics[width=3.2in]{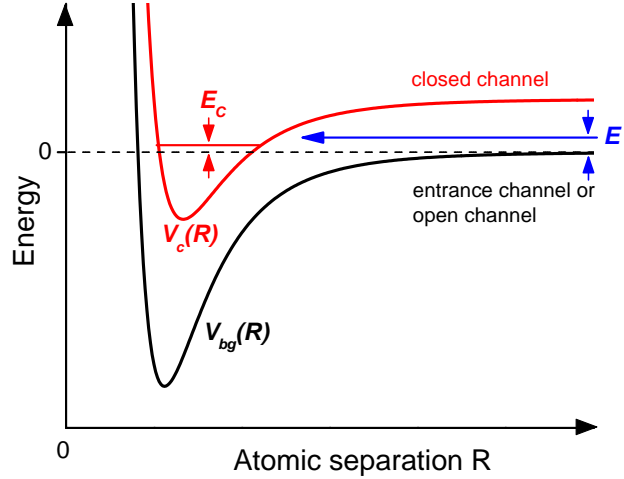}
\caption{Basic two-channel model for a Feshbach resonance. The
phenomenon occurs when two atoms colliding at energy $E$ in the entrance channel resonantly couple to a molecular bound state with energy
$E_c$ supported by the closed channel potential. In the ultracold domain,
collisions take place near zero-energy, $E \rightarrow
0$.  Resonant coupling
is then conveniently realized by magnetically tuning $E_c$ near 0  if the
magnetic moments of the closed and open channel differ.} \label{fig:intro_potential}
\end{figure}

In a simple picture, we consider two molecular potential curves $V_\mathrm{bg}(R)$ and $V_\mathrm{c}(R)$, as illustrated in Fig.~\ref{fig:intro_potential}.
For large internuclear distances $R$, the background potential
$V_\mathrm{bg}(R)$ asymptotically correlates with two free atoms in an
ultracold gas. For a collision process with a very small energy
$E$, this potential represents the energetically open channel, usually referred to as the {\em entrance channel}. The other potential, $V_\mathrm{c}(R)$, representing the {\em closed channel},
is important as it can support bound molecular states near the
threshold of the open channel.

A Feshbach resonance occurs when the bound molecular state in the
closed channel is energetically close to the scattering state in
the open channel, and some coupling leads to a mixing
between the two channels. The energy difference can be controlled
via a magnetic field when the corresponding magnetic moments are
different. This leads to a {\em magnetically tunable Feshbach
resonance}, which is the common way to achieve
resonant coupling in the experiments with ultracold gases where the collision energy is practically zero.

\begin{figure}
\includegraphics[width=3.2in]{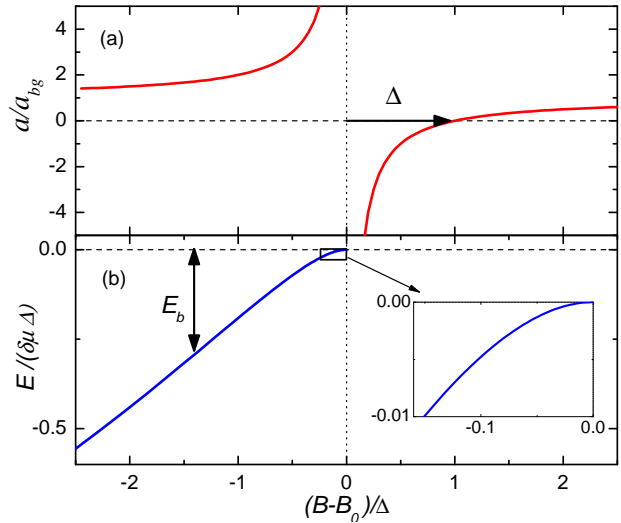}
\caption{Scattering length $a$ (upper panel) and molecular state energy $E$ (lower panel) near a magnetically tuned Feshbach resonance.
The inset shows the universal regime near the point of resonance where $a$ is very large and positive.} \label{fig:intro}
\end{figure}

A magnetically tuned Feshbach resonance can be described by a simple
expression  for the $s$-wave scattering length $a$ as a function of the magnetic field $B$,
\begin{equation}
 a(B) = a_\mathrm{bg} \left( 1 - \frac{\Delta}{B-B_0} \right) \,.
      \label{eq:sclength}
\end{equation}
Figure~\ref{fig:intro}(a) illustrates this resonance expression.
The {\em background scattering length}
$a_\mathrm{bg}$, which is the scattering
length of $V_\mathrm{bg}(R)$, represents the off-resonant value.  It is directly related to the energy of the last bound vibrational level of $V_\mathrm{bg}(R)$. The parameter $B_0$ denotes the {\em resonance
position}, where the scattering length diverges
($a\rightarrow\pm\infty$), and the parameter $\Delta$ is
the {\em resonance width}.

The energy of the weakly bound molecular state near the resonance
position $B_0$  is shown in Fig.~\ref{fig:intro}(b), relative to the threshold of two free atoms with zero kinetic energy. The energy approaches the threshold on
the side of the resonance where $a$ is large and positive.   Away from the resonance, the
energy varies linearly with $B$ with a slope given by $\delta \mu$, the difference in magnetic moments of the open and closed channels.  Near
the resonance the coupling between the two channels mixes in entrance-channel contributions and strongly bends the molecular state.

In the vicinity of the resonance position at $B_0$, where the two
channels are strongly coupled, the scattering length is very
large. For large positive $a$, a ``dressed'' molecular state exists  with a binding energy given by
\begin{equation}
 E_{\rm b} = \frac{\hbar^2}{2 m_{\rm r} a^2}\,,
 \label{Eb}
\end{equation}
where $m_{\rm r}$ is the reduced mass of the atom pair.  In this limit, $E_{\rm b}$ depends quadratically on the magnetic detuning $B-B_0$ and results in the bend depicted in the inset of Fig.~\ref{fig:intro}.  This region is of particular interest because of its {\em universal} properties; see discussions in Chapters 6 and 11. Molecules formed in this regime are extremely weakly bound, and they are described by a wavefunction that extends far out of the classically allowed range. Such exotic molecules are therefore commonly referred to as {\em halo dimers}; we shall discuss their intriguing physics in Sec.\ \ref{sec:halo}.

These considerations show how a Feshbach resonance is inherently connected with a weakly bound molecular state. The key question for experimental applications is how to prepare the molecular state in a controlled way; we shall address this in Sec.\ \ref{sec:makedetect}.

\subsection{Binding energy regimes}
\label{ssec:term}

The name {\em Feshbach molecule} emphasizes the production method, as it commonly refers to diatomic molecules made in ultracold atomic gases via Feshbach resonances.
But what are the physical properties associated with a Feshbach molecule, and what is the exact definition of such a molecule? It is quite obvious that a Feshbach molecule is a highly excited molecule, existing near the dissociation threshold and having an extremely small binding energy as compared to the one of the vibrational ground state. To give an exact definition, however, is hardly possible as the properties of molecules gradually change with increasing binding energy, and there is no distinct physical property associated with a Feshbach molecule.
Molecules created via Feshbach resonances can be transferred to many other states near threshold (Sec.~\ref{sec:manipulation}) or to much more deeply bound states (Sec.~\ref{sec:towardground}), thus being converted to more conventional molecules. There is no really meaningful definition of when a molecule can still be called a Feshbach molecule, or when it loses this character. We therefore use a loose definition and call a Feshbach molecule any molecule that exists near the threshold in the energy range set by roughly the quantum of vibrational energy. Any other molecule we consider a deeply bound molecule, although views may differ on this.

\begin{figure}
\includegraphics[width=3.5in]{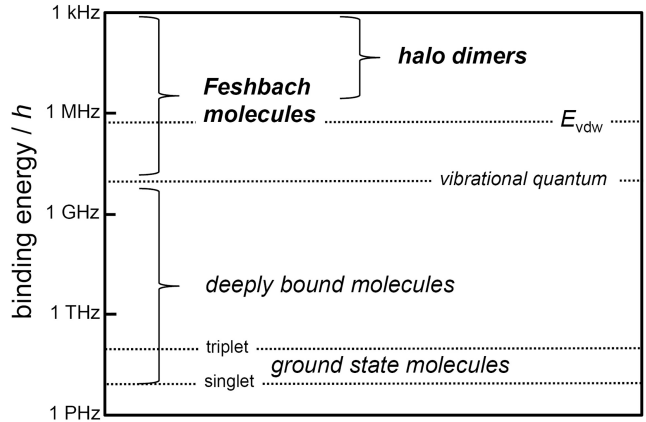}
\caption{Binding energy regimes for the example of Cs$_2$ dimers, illustrated on a logarithmic scale. For the long-range van der Waals attraction between the atoms, a characteristic energy $E_{\rm vdw}$ is introduced; see Chapter~6 and Sec.~\ref{sec:halo}. The vibrational quantum, here defined as the energy range below threshold in which one finds at least one bound state, corresponds to about $40\,E_{\rm vdw}$; see Fig.~4 in Chapter~6. The binding energies of the vibrational ground state of the triplet and the single potential are typically five or six orders of magnitudes larger than this vibrational quantum.}
\label{fig:energyscales}
\end{figure}

Figure \ref{fig:energyscales} illustrates the vast range of molecular binding energies for the example of Cs$_2$ dimers. According to our definition Feshbach molecules may have binding energies of up to roughly $h \times 100$\,MHz for the heavy Cs$_2$ dimer and up to $h \times 2.5$\,GHz for the light Li$_2$ dimer. These values are very small compared to the binding energies of ground state molecules, which are of the order of $h \times 10$\,THz or $h \times 100$\,THz for the triplet and singlet potential, respectively. Nevertheless, ultracold Feshbach molecules can serve as the starting point for state transfer to produce very deeply bound, and even ground-state molecules, as we shall discuss in Sec.~\ref{sec:towardground}.

Two particular regimes of Feshbach molecules can be realized close to the dissociation threshold. The {\em halo regime} requires binding energies well below the so-called van der Waals energy $E_{\rm vdw}$ (Sec.\ II in Chapter 6), which corresponds to about 3\,MHz for Cs$_2$ and about 600\,MHz for Li$_2$. Molecules in high partial-wave states (Sec.~\ref{sec:manipulation}) can exist above the dissociation threshold, as the large centrifugal barrier prevents them from decaying; such {\em metastable Feshbach molecules} have a negative binding energy.

\section{Making and detecting Feshbach molecules}
\label{sec:makedetect}

\begin{figure}
\includegraphics [width=5in]{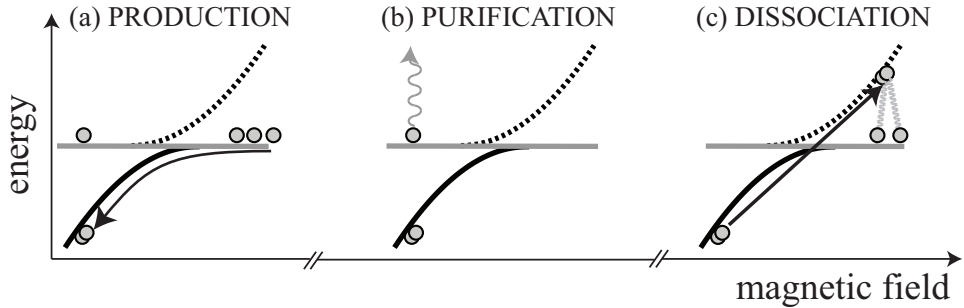}
\caption{Illustration of a typical experimental sequence to create, purify, and detect a sample of Feshbach molecules, (a) association via a magnetic-field sweep across the resonance, (b) selective removal of the remaining atoms, and (c) forced dissociation through a reverse magnetic field sweep. The
dissociation is usually followed by imaging of the resulting cloud of atoms. The solid line corresponds to the bound molecular state, which intersects the threshold (gray horizontal line) and causes the Feshbach resonance; see also Fig.~\ref{fig:intro}. The dashed line indicates the molecular state above threshold, where it has the character of a quasi-bound state coupling to the scattering continuum.}
\label{fig:ProfPurfDetec}
\end{figure}

\begin{figure}
\includegraphics[width=2.7in]{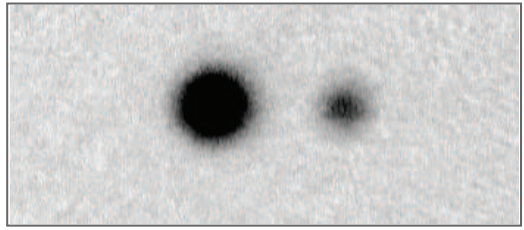}
\caption{Example for the preparation of a pure molecular cloud. Here an optically trapped BEC of $^{87}$Rb atoms was subjected to a magnetic-field sweep across a Feshbach resonance. The Stern-Gerlach technique was then applied after release from the trap to spatially separate the atoms (left) from the molecules (right). Finally the molecules were dissociated by a reverse Feshbach ramp, and an absorption image was taken. The field of view is 1.7\,mm $\times$ 0.7\,mm. Courtesy of S.~D\"urr and G.~Rempe.} \label{fig:SGabsorption}
\end{figure}

This Section discusses how to create  ultracold molecules, starting from a
degenerate or near-degenerate atomic gas. The three main experimental steps are illustrated in Fig.\,\ref{fig:ProfPurfDetec}. In a first step (a) atoms are converted into molecules, which can be done by sweeping the magnetic field across a Feshbach resonance. In a second step (b) a purification scheme can be applied that removes all the remaining atoms. In a third step (c) the molecules are forced to dissociate in a reverse sweep across the resonance in order to detect the resulting atoms.
As an example, Figure \ref{fig:SGabsorption} shows an image taken after the Stern-Gerlach separation of an atom-dimer mixture released from a trap; the smaller cloud represents a pure sample of Feshbach molecules. Similar techniques can also be applied to trapped atoms. In the following we describe the three steps of production, purification, and dissociation after a discussion of the important role of quantum statistics.

\subsection{Bosons and fermions: role of quantum statistics}
\label{ssec:statistic}

Ultracold diatomic molecules can be composed either of two bosons, two fermions, or a boson and a fermion. In the first two cases, the molecules have  bosonic character, while in the third case the molecules are
fermions. Bosons are described by wavefunctions, which are symmetric with respect to exchange of identical particles, while, in the fermionic case, the
wavefunctions are antisymmetric. This fundamental difference has important
consequences.


Identical bosons (fermions) can collide in even (odd) partial waves. The role of partial waves is discussed in Chapter 1. Here we use  $\ell$ as the corresponding angular momentum quantum number. The first partial-wave contribution for bosons (fermions) is then the
$s$-wave with $\ell=0$ ($p$-wave wit $\ell=1$) \cite{notepw}. At ultralow temperatures,  only $s$-wave collisions are dominant with the consequence that collisions between identical fermions are suppressed. The absence of $s$-wave collisions is also the reason why, for instance, direct evaporative cooling can not be applied to a sample of identical fermions. The situation changes when fermionic atoms are in
different internal states, like hyperfine or Zeeman sublevels. The distinguishable particles can then interact in any partial wave. Fermion-composed molecules in $s$-wave states are therefore generally associated from ultracold two-component spin mixtures. Interactions in all partial waves are obviously allowed if two different atomic species are involved, regardless of their fermionic or bosonic character.

The quantum-statistical character of a molecule can crucially influence its collisional stability. Feshbach molecules in highly excited vibrational states are in principle very sensitive to vibrational relaxation induced by atom-molecule and molecule-molecule collisions. If such an inelastic relaxation process occurs, the energy release is very large as compared to the trap depth and will result in an immediate loss of all collision partners from the trap.  A remarkable exception to this general behavior are Feshbach molecules composed of fermionic atoms in a halo state, as they exhibit an extremely high stability against inelastic decay. The reason of this different behavior is a
Pauli suppression effect. In a two-component Fermi mixture, relaxation processes are unlikely since they necessarily involve at least two identical fermions; for more details see Chapter~10 and
Ref.~\cite{Petrov2004wbm}. An intermediate case can be found for molecules composed in a Bose-Fermi mixture of two atomic species. Here, the collisional stability depends on the atomic partner involved in a collision,
either a boson or a fermion. Only in collisions with fermionic atoms, a similar Pauli suppression effect can be expected

The different degree of stability of bosons and fermions leads to a surprising
result. One might think that the best way to achieve mBEC is the direct conversion of an atomic BEC. However, the stability of fermions turned out to be a key ingredient to experimentally achieve the condensation of molecules; see Sec.\,\ref{ssec:fermi}.  For attaining a collisionally stable mBEC of boson-composed molecules, their transfer into the rovibrational ground state (Sec.~\ref{sec:towardground}) may be the only feasible way.


\subsection{Overview of association methods}
\label{ssec:assoc}

The optimum association strategy depends on several factors
connected to the specific system under investigation. The main factors to
consider are the quantum-statistical nature of the atoms that form the molecule and the particular Feshbach resonance employed. Ultralow temperatures and high phase-space
densities are essential requirements for an efficient molecule creation.
Therefore all association techniques start with an ultracold trapped atomic
sample.

The most widespread  technique for molecule association uses time-varying
magnetic fields. The magnetic field modifies the energy difference between the
scattering state and the molecular state. When the two states have the same energy, the
scattering length diverges and a molecular state can be accessed in the region
of positive scattering length ($a>0$); see Sec.~\ref{ssec:feshres}.

Magnetic field time-sequences include linear ramps, fast jumps, or oscillating
magnetic fields.  Figure \ref{fig:ProfPurfDetec}(a) illustrates a
magnetic-field ramp, commonly referred to as ``Feshbach ramp''
\cite{vanAbeelen1999tdf,Mies2000mof,Timmermans1999fri}.  A homogeneous magnetic field is swept across a Feshbach resonance from the side where $a<0$ to the side where $a>0$. The coupling between atom pairs and molecules removes
the degeneracy at the crossing of the two corresponding energy levels. The resulting avoided level crossing can be used to adiabatically convert atom pairs into molecules by sweeping the magnetic field across the resonance, as indicated by the arrow in Fig.\,\ref{fig:ProfPurfDetec}(a). The atomic sample is partially converted into a molecular gas.
In the fermionic case, the conversion
efficiency does solely depend  on the ramp speed and on the atomic phase-space
density \cite{Hodby2005peo}, and can reach values close to one. In the case of
bosons, the conversion efficiency is affected  by inelastic collision
processes. Near the center of a Feshbach resonance, the atoms experience a huge increase of the three-body recombination rate, while, after the association,
the atom-dimer mixture can undergo fast collisional relaxation. The best
parameters for molecule association require a compromise for the optimum ramp
speed, which should be slow enough for an efficient conversion, but fast enough to avoid detrimental losses.

Inelastic collisional loss can be reduced by using more elaborate time sequences for the magnetic field or, alternatively, resonant radio-frequency techniques. The key idea is to introduce an efficient atom-molecule coupling while minimizing the time spent near the resonance, where strong loss and heating results from inelastic interactions. An efficient technique is to apply a small sinusoidal modulation to the homogeneous field. The oscillating field then induces a corresponding modulation of the energy difference between the scattering state and the molecular state. The molecules are associated when the modulation
frequencies match the molecular binding energies \cite{Hanna2007aom}. Typical
modulation frequencies range from a few kHz to a few 100 kHz.  The signature of molecule production is a resonant decrease of the atom number, as observed in
gases of $^{85}$Rb \cite{Thompson2005ump} and $^{40}$K \cite{Gaebler2007pwf}, and in a mixture of the two isotopes $^{85}$Rb and $^{87}$Rb \cite{Papp2006ooh}. A resonant coupling between atom pairs and molecules can also be induced by a radio-frequency excitation, which stimulates a transition between the atom pair state and the molecular state without modulating the energy difference. Heteronuclear $^{40}$K$^{87}$Rb molecules were produced with this technique both in an optical lattice \cite{Ospelkaus2006uhm} and in an optical dipole trap \cite{Zirbel2008cso}.

Collisional losses can also be suppressed by using a
suitable trapping environment. This is the case when atoms are prepared in an
optical lattice with double site occupancy. A stable system with a single
molecule in each lattice site can then be created using, for instance, a
Feshbach ramp \cite{Thalhammer2006llf, Volz2006pqs}.

\begin{figure}
\includegraphics[width=2.8in]{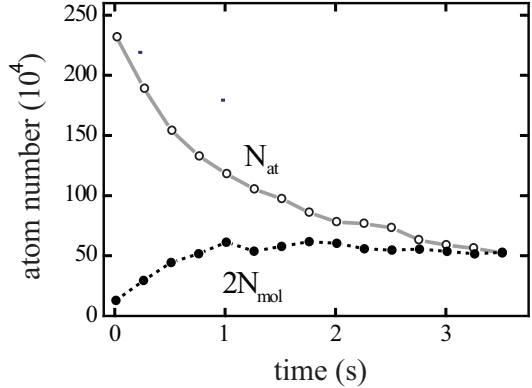}
\caption{Formation of $^6$Li$_2$ halo dimers via three-body recombination. The molecules are created in an optically trapped spin mixture of atomic $^6$Li at a temperature of 3\,$\mu$K. The experiment is performed near a broad Feshbach resonance, where $a=+1420 a_0$ and
$E_{\text b}=k_{\text B}\times 15\,\mu{\text K}=h\times 310$\,kHz. $N_{\rm at}$ and $N_{\rm mol}$ denote the number of unbound atoms and the
number of molecules, respectively. Adapted from Ref.~\cite{Jochim2003pgo}.}
\label{fig:threebodytherm}
\end{figure}

A completely different association strategy can be applied to create molecules in an atomic Fermi gas of $^6$Li atoms. This special situation is particularly important as it has opened up a simple and efficient route towards mBEC \cite{Jochim2003bec, Zwierlein2003oob}; see also Sec.~\ref{ssec:fermi}. Near a Feshbach resonance, in the regime of very large positive scattering lengths, atomic {\em three-body recombination} efficiently produces halo dimers. In a three-body recombination event in a two-component atomic Fermi gas, two non-identical fermions bind into a molecule while the third atom carries away the leftover energy and momentum. Usually, the energy released in three-body collisions is very large so that all collision partners immediately leave the trap. However, for the special case of halo dimers, the very small binding energies can be on the order of the typical thermal energies of the trapped particles and thus less than the trap depth \cite{Varenna2006}. The formation of halo dimers via three-body recombination then proceeds with small heating and negligible trap losses. This way of forming molecules can also be
interpreted in terms of a thermal atom-molecule equilibrium
\cite{Chin2004tea,Kokkelmans2004dam}. Typical time scales for the atom-molecule thermalization are on the order of 100~ms to 1~s, requiring the high collisional stability of fermion-composed molecules in the halo regime.
As an example, Fig.\,\ref{fig:threebodytherm} shows how an initially pure sample of $^6$Li atoms changes into an atom-molecule mixture \cite{Jochim2003pgo}.

\subsection{Purification schemes}
\label{ssec:purify}

Purification is an important step in many experiments with Feshbach molecules. The removal of atoms can be needed to avoid fast collisional losses of molecules. In other cases,
pure molecular samples are needed for particular applications. Most of the
purification schemes  act selectively  on atoms and they are applied on time scales short compared to the typical molecular lifetimes.

A possible purification strategy exploits the difference in magnetic moments of molecules and atoms. The two components of the gas can be spatially separated with the Stern-Gerlach technique, which uses magnetic field gradients
\cite{Herbig2003poa, Durr2004oom}; for an example see Fig.\,\ref{fig:SGabsorption}. This method is usually applied to gases in free space after release from the trap.

A fast and efficient purification method, well suited for the application to a trapped cloud, uses resonant light to push the atoms out of the sample by resonant radiation pressure \cite{Xu2003foq, Thalhammer2006llf, Mark2007sou}. The ``blast'' light needs to be resonant with a cycling atomic transition, which allows for the repeated scattering of photons. In many cases, the atoms are not in a ground-state sublevel connected to a cycling optical transition, so that an intermediate step is necessary. In the most simple implementation, this is an optical pumping step that selectively acts on the atoms without affecting the molecules.
Alternatively a microwave pulse can be used to transfer the atoms into the appropriate atomic level. The high selectivity of this double-resonance method minimizes residual heating and losses of molecules from the trap.

\subsection{Detection methods}
\label{ssec:detect}

A very efficient way to detect Feshbach molecules is to convert them back into
atom pairs and to take an image of the reconverted atoms with standard
absorption imaging techniques \cite{Herbig2003poa,Durr2004oom}.  In principle, each of the above-described association methods can be reverted and turned into a corresponding dissociation scheme.

A robust and commonly used dissociation scheme is to simply reverse the Feshbach ramp, as illustrated in Fig.\,\ref{fig:ProfPurfDetec}(c). The reverted ramp is usually
 chosen to be much faster than the one employed for association.
 The molecules  are then brought into a quasi-bound state above the atomic threshold. Here they quickly dissociate
into atom pairs, converting the dissociation energy into kinetic energy. The
back-ramp is usually done in free space and can be performed either immediately
after release from the trap or after some expansion time ($\approx 10-30$ ms).
The choice of the delay time between the dissociation and the detection sets the type
of information that can be extracted from the absorption image. For a long delay
time, the sample expands thus mapping the information on the momentum distribution
onto the spatial distribution. The image then contains quantitative information on the dissociation energy \cite{Mukaiyama2004dad}. For a short delay time, the absorption image simply reflects the
spatial distribution of the molecules before the dissociation.

Fast and efficient dissociation requires a sufficiently strong coupling of the
molecular states  to the scattering continuum. Molecular states with high
rotational quantum number $\ell$  typically have weak couplings and high
centrifugal barriers.  A striking example is provided by Cs$_2$ Feshbach
molecules with  $\ell=8$  \cite{notepw}. These molecules do not sufficiently
couple to the atomic continuum and  dissociation is prevented. There are two
ways to detect them. The first is based on a time reversal of the association path (Sec.\,\ref{ssec:cruising}) \cite{Mark2007siw}. The second exploits the crossing and the mixing with a quasi-bound state above threshold with $\ell\leq 4$ to force the dissociation \cite{Knoop2008mfm}.

\begin{figure}
\includegraphics[width=3.5in]{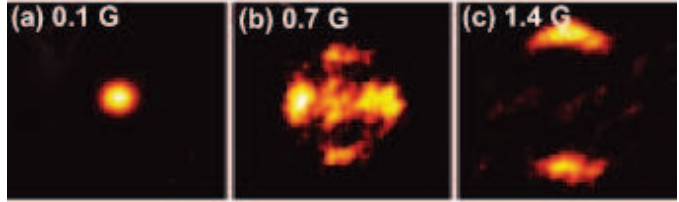}
\caption{Dissociation patterns of $^{87}$Rb$_2$ molecules near a $d$-wave shape
resonance. The molecules can dissociate either directly into the continuum or
indirectly by passing through a $d$-wave shape-resonance state located behind
the centrifugal barrier. Initially, the direct dissociation process dominates
and preferentially populates isotropic $s$-wave states (a). Approaching the shape resonance, the pattern reveals first an interference between $s$- and $d$-partial waves (b), and then shows a pure outgoing $d$-wave (c). The given magnetic field values are relative to the field where dissociation is observed to set in. Adapted from Ref.~\cite{Volz2005fso}.} \label{fig:dissoc}
\end{figure}

Dissociation patterns can also provide additional spectroscopic information. An example is shown in Fig.\,\ref{fig:dissoc}. Here the dissociation patterns of
$^{87}$Rb$_2$ show strong modifications by the presence of a $d$-wave shape
resonance \cite{Volz2005fso}.

In the halo regime, molecules can be detected by direct imaging \cite{Zwierlein2003oob, Bartenstein2004cfa}. The imaging
frequency for halo molecules is very close to the atomic frequency. Presumably, the first photon dissociates the molecule into two atoms, which then absorb the subsequent photons. In the case of heteronuclear molecules, directed imaging is possible in a wider range of binding energies. This happens because  the
excited and ground molecular potentials both vary as $1/R^6$; see Chapter 5. In the homonuclear case, the excited potential varies as $1/R^3$ \cite{Zirbel2008hmi}.

\section{Internal-state manipulation near threshold}
\label{sec:manipulation}

Below the atomic threshold a manifold of molecular states exists.
They correspond to different vibrational, rotational and magnetic
quantum numbers, and belong to potentials correlated with different
hyperfine states. Having different magnetic moments, many levels
cross when the magnetic field is varied. When some coupling between
two molecular states that cross is present an avoided crossing
occurs. The concept of avoided level crossings is very general and
has found important applications in many fields of physics. We have
already discussed the application to associate molecules in
Sec.~\ref{sec:makedetect}.

In the field of ultracold molecules, the information on the avoided
crossings is fundamentally important to fully describe the molecular
spectrum and to understand the coupling mechanisms between different
states. At the same time controlled state transfer near avoided
crossings opens up the intriguing possibility to populate many
different molecular states that are not directly accessible by
Feshbach association.

\subsection{Avoided level crossings}
\label{ssec:crossing}

\begin{figure}
\includegraphics[width=3in]{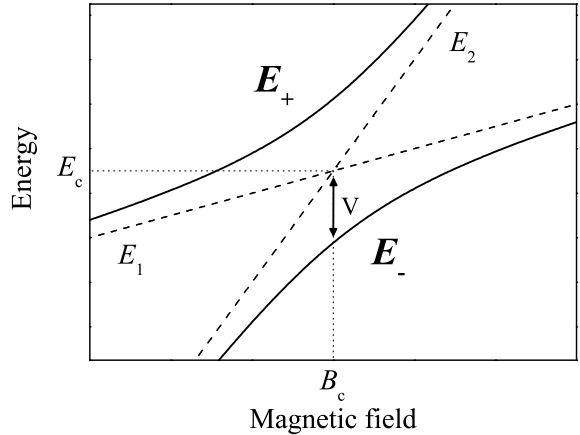}
\caption{Schematic representation of an avoided crossing between two
molecular states. The dashed lines represent the diabatic states
with energies $E_1$ and $E_2$, with a crossing at $B_{\rm c}$. The
solid lines represent the adiabatic states with energies $E_+$ and
$E_-$, caused by a coupling $V$ between the two diabatic states.}
\label{fig:avoidedcrossing}
\end{figure}

If an avoided crossing is well separated from any other avoided
crossing, including the one resulting near threshold from the
coupling to the scattering continuum, a simple two-state model can
be applied. Consider two molecular states ($i=1,2$), for which the
magnetic field dependencies of the binding energies $E_i$ of the
diabatic states $\varphi_i$ are given by
\begin{equation}
E_i(B)=\mu_i(B-B_{\rm c})+E_{\rm c}, \\
\end{equation}
where $\mu_i$ are the magnetic moments, and $B_{\rm c}$ and $E_{\rm
c}$, the magnetic field and energy at which the two states cross,
respectively; see Fig.~\ref{fig:avoidedcrossing}. The crossing is
conveniently described in terms of adiabatic states if the two
states are coupled by the interaction Hamiltonian $H$, i.e. $\langle
\varphi_1 | H | \varphi_2 \rangle \neq 0$. To obtain the
corresponding adiabatic energies one has to solve the following
eigenvalue problem,
\[ \left(\begin{array}{cc}
E_1 & V \\
V & E_2 \\
\end{array}
\right) \left(\begin{array}{c}
\varphi_1 \\
\varphi_2 \\
\end{array}
\right)=E\left(\begin{array}{c}
\varphi_1 \\
\varphi_2 \\
\end{array}
\right),
\]
where $V=\langle \varphi_1 | H | \varphi_2 \rangle$ is the coupling
strength between the diabatic states. The adiabatic energies are
given by
\begin{equation}
2E_{\pm}=(E_1+E_2)\pm\sqrt{(E_1-E_2)^2+4V^2},
\end{equation}
where the energies $E_+$ and $E_-$ correspond to the upper and lower
adiabatic levels of the avoided crossing. The energy difference
between the adiabatic levels is
\begin{equation}\label{eq:DeltaE} \Delta
E=|E_+-E_-|=\sqrt{(\mu_1-\mu_2)^2(B-B_{\rm c})^2+4V^2},
\end{equation}
which at the crossing is twice the coupling strength, i.e. $\Delta
E(B_{\rm c})=2V$.

\begin{figure}
\includegraphics[width=4in]{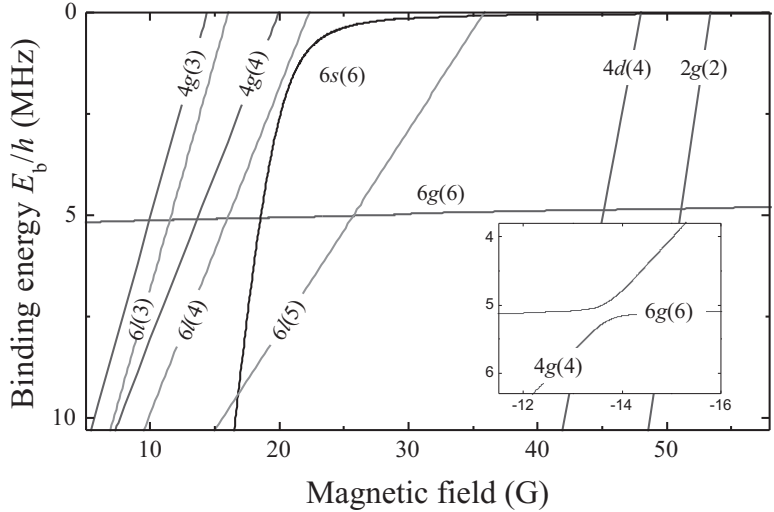}
\caption{Molecular energy structure of Cs$_2$ below the threshold of
two free Cs atoms in the absolute hyperfine ground-state sublevel.
The molecular states near threshold are well characterized by
quantum numbers $|f,m_f;\ell,m_\ell\rangle$ \cite{Kohler2006poc},
where $f$ represents the sum of the total atomic spins $F_{1,2}$ of
the individual atoms, and $\ell$ is the rotational quantum number.
The respective projection quantum numbers are given by $m_f$ and
$m_\ell$. The interaction Hamiltonian conserves $f+\ell$ at zero
magnetic field, and always conserves $m_f+m_\ell$. In this example,
Feshbach association is performed with pair of atoms that has
$m_f+m_\ell=6$, and only molecular states with $m_f+m_\ell=6$ have
to be considered. Therefore the labeling $f \ell (m_f)$ is
sufficient to characterize the molecular states. Due to relatively
strong relativistic spin-spin dipole and second-order spin-orbit
interactions, molecular states with rotational quantum number up to
$l$-wave ($\ell=8$) have to be taken into account in the case of Cs.
The inset shows one of the narrow avoided crossing between two
molecular states. Adapted from Ref.~\cite{Mark2007sou}.}
\label{fig:molspectrumCs2}
\end{figure}

Let us illustrate the occurrence of avoided level crossings by the
example of Cs$_2$, for which the near-threshold spectrum is shown in
Fig.~\ref{fig:molspectrumCs2}. The molecular spectrum of Cs$_2$
provides many avoided crossings, basically for two reasons. First of
all, there exists a high density of molecular levels because its
large mass gives rise to a small vibrational and rotation spacing,
and its large nuclear spin gives rise to many hyperfine substates.
Secondly, relatively strong relativistic spin-spin dipole and
second-order spin-orbit interactions couple many of these states.
Most crossings in Fig.~\ref{fig:molspectrumCs2} are indeed avoided
crossings and the inset shows one in an expanded view as an example.
As they are well separated from each other, the simple two-state
model applies very well.

\begin{figure}
\includegraphics[width=5.6in]{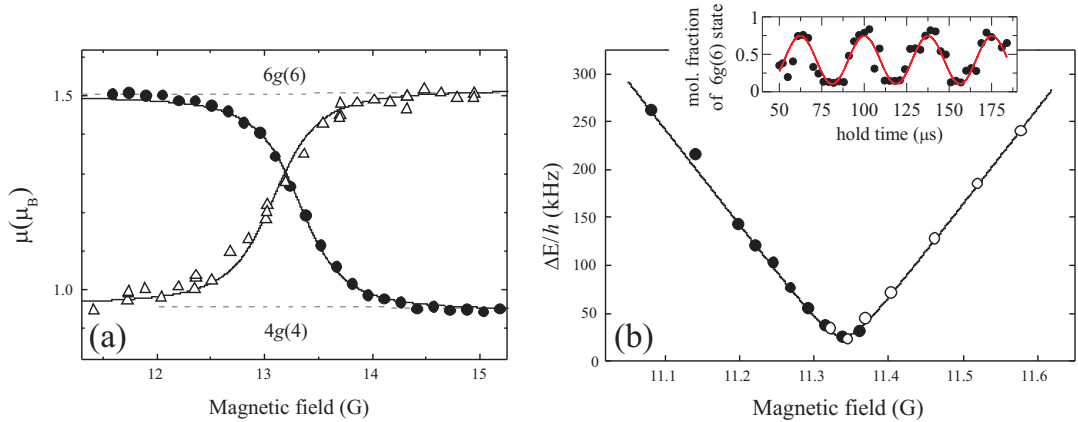}
\caption{Examples of two different techniques to map avoided
crossings, both applied to Cs$_2$. (a) Magnetic moment spectroscopy
on the two molecular states around the $6g(6)$/$4g(4)$ avoided
crossing, measuring the magnetic moment as function of the magnetic
field. A fit to the data gives $B_{\rm c}$=13.29(4)\,G and
$V=h\times164(30)$\,kHz. (b) St\"{u}ckelberg interferometry on the
two molecular states around the $6g(6)$/$6l(3)$ avoided crossing.
The solid curve is a fit according to Eq.~\ref{eq:DeltaE}, resulting
in $B_{\rm c}$=11.339(1)\,G and $V=h\times14(1)$\,kHz. Inset: raw
data showing an oscillation of the population in state $6g(6)$ right
in the center of the crossing. Panel (a) adapted from
Ref.~\cite{Mark2007sou}, panel (b) from Ref.~\cite{Mark2007siw}.}
\label{fig:avoidedcrossing_exp}
\end{figure}

The crossing position $B_{\rm c}$ and the coupling strength $V$ can
be measured by different methods. Magnetic moment spectroscopy
relies on measuring the effect of a magnetic field gradient after
the molecular cloud is released from the trap. The difference in the
position of a molecular cloud after a fixed time-of-flight with and
without a magnetic field gradient is directly proportional to the
magnetic moment. In the avoided crossing region the magnetic moment
of the adiabatic molecular states shows a rapid change with magnetic
field. This technique has been applied to map out several avoided
crossings of the Cs$_2$ spectrum, from which one is shown in
Fig.~\ref{fig:avoidedcrossing_exp}(a). Another method relies on
direct probing of $\Delta E$ by radio-frequency excitation between
the adiabatic states, as demonstrated for $^{87}$Rb$_2$
\cite{Lang2008ctm}.

More sophisticated methods rely on Ramsey-type or St\"{u}ckelberg
interferometry, which have been applied to avoided crossings in
$^{87}$Rb$_2$ \cite{Lang2008ctm} and Cs$_2$ \cite{Mark2007siw},
respectively. In the Ramsey scheme a radio-frequency pulse creates a
coherent superposition. After a hold time and a second pulse, the
population in one of the molecular branches is measured. This
population shows an oscillation as function of the hold time with a
frequency of $\Delta E/h$. In the St\"{u}ckelberg scheme a coherent
superposition is obtained by a magnetic field sweep over the avoided
crossing; see Sec.~\ref{ssec:cruising}. After a hold time, a reverse
magnetic field sweep is applied and the population in both molecular
branches is measured. Here the oscillation frequency in the
population is also equal to $\Delta E/h$. In
Fig.~\ref{fig:avoidedcrossing_exp}(b) the mapping of a very narrow
avoided crossing with the St\"{u}ckelberg scheme is shown.

\subsection{Cruising through the molecular spectrum}
\label{ssec:cruising}

A magnetic field ramp over the avoided crossing can be used for
state-transfer. When the magnetic field is ramped slowly, the
population follows the adiabatic states. This is a called an
adiabatic ramp. If the change in magnetic field is very fast, the
molecules do not experience the coupling between the diabatic
states, and therefore remain in their initial state. This is a
non-adiabatic ramp. The well-known Landau-Zener model describes the
final population in both molecular states after the ramp. The
probability of adiabatic transfer is given by
\begin{equation}\label{eq:LZ}
p=1-\exp\left(-\dot{B_{\rm c}}/|\dot{B}|\right),
\end{equation}
where $\dot{B}$ is the linear ramp speed and $\dot{B_{\rm c}}=2\pi
V^2/(\hbar \left|\mu_1-\mu_2\right|)$. An adiabatic ramp corresponds
to a ramp speed of $|\dot{B}|\ll \dot{B_{\rm c}}$, while a
non-adiabatic ramp requires $|\dot{B}|\gg \dot{B_{\rm c}}$.

Manipulation of the population over avoided crossings is fully
coherent as there is no dissipation. Intermediate ramp speeds result
in a coherent splitting of the population over the two molecular
states. This property was demonstrated by a St\"{u}ckelberg
interferometer, in which two subsequent passages through an avoided
crossing in combination with a variable hold time lead to
high-contrast population oscillations, as shown in the inset of
Fig.~\ref{fig:avoidedcrossing_exp}(b). The required ramp speed for
equal splitting is on the order of 1\,G/$\mu$s for an avoided
crossing with $V\sim h\times100$\,kHz.

\begin{figure}
\includegraphics[width=2.8in]{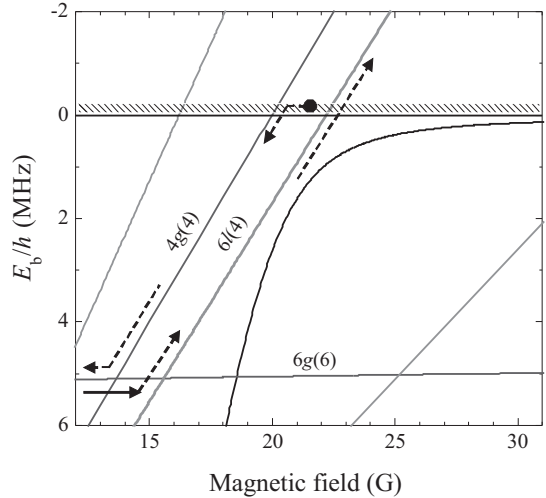}
\caption{Population scheme for the Cs$_2$ $6l(4)$ state. Starting
with Feshbach association on the 19.8-G $g$-wave Feshbach resonance,
the resulting $4g(4)$ Feshbach molecules are brought to larger
binding energies by ramping down the magnetic field. At a binding
energy of about $h\times5$\,MHz, the avoided crossing with the
$6g(6)$ state is followed adiabatically. Then a fast ramp in the
opposite direction is made, jumping the avoided crossing, after
which a slow magnetic field ramp towards higher magnetic fields
leads to adiabatic transfer from the $6g(6)$ to the $6l(4)$ state.
The molecular sample in the $6l(4)$ state can be brought above the
atomic threshold, as the large centrifugal barrier prohibits
dissociation. Adapted from Ref.~\cite{Knoop2008mfm}.}
\label{fig:metastablepopulation}
\end{figure}

The magnetic field ramps over avoided crossings allow to populate
states that are not directly accessible by Feshbach association.
This includes states that do not cross the atomic threshold, for
example the $6g(6)$ state in Fig.~\ref{fig:molspectrumCs2}. There
are also states that do cross the atomic threshold, but do not
induce Feshbach resonances as the coupling with the atomic continuum
is too weak. This occurs for molecular states with high rotational
quantum number. For Cs$_2$ molecular states with $\ell>4$, e.g.\ the
$l$-wave states shown in Fig.~\ref{fig:molspectrumCs2}, no Feshbach
resonances exist. In Refs.~\cite{Mark2007sou,Knoop2008mfm} the
population of the $l$-wave states using avoided crossings was
demonstrated. As an example, the population scheme for one of the
$l$-wave states is shown in Fig.~\ref{fig:metastablepopulation}.

The ability to populate molecular states that do not induce Feshbach
resonances raises the question on the lifetime of these states above
the atomic threshold, as the vanishing coupling with the atomic
continuum implies that dissociation is suppressed. By studying the
lifetime of the molecular sample above threshold it was found that
$l$-wave molecules are indeed stable against dissociation on a
timescale of 1\,s \cite{Knoop2008mfm}. The preparation of long-lived
Feshbach molecules above the atomic threshold opens up the
possibility to create novel metastable quantum states with strong
pair correlations.

%

The method of jumping avoided crossings is in practice limited to
crossings with energy splittings up to typically $h\times200$\,kHz.
For stronger crossings, fast ramps are technically impracticable and
the crossings are followed adiabatically when the magnetic field is
ramped, leaving no choice for the state transfer. This problem can
be overcome with the help of radio-frequency excitation
\cite{Lang2008ctm}. For $^{87}$Rb$_2$ the transfer of molecules over
nine level crossings was demonstrated when the magnetic field was
ramped down from a Feshbach resonance near 1~kG to zero. This
efficiently produced molecules with a binding energy $E_{\rm b} = h
\times 3.6$\,GHz at zero magnetic field. The combination of these
elaborate ramp and radio-frequency techniques allows for cruising
through the complex molecular energy structure. The level spectrum
serves as a molecular ``street map'' and by passing straight through
crossings or performing left or right turns one can reach any
desired destination.

\section{Halo dimers}\label{sec:halo}

Very close to resonance the Feshbach molecules are extremely weakly
bound and become \emph{halo dimers}. This halo regime is interesting
because of its universal properties, regarding its intrinsic
behavior (Sec.~\ref{ssec:universal}) and its few-body interaction
properties (Sec.~\ref{ssec:halocoll}), in particular in the context
of Efimov physics (Sec.~\ref{ssec:efimov}). The properties of
fermion-composed halo dimers, extensively discussed in Chapter 10,
allow the realization of molecular Bose-Einstein condensation
(mBEC), which is the topic of Sec.~\ref{ssec:fermi}.

\subsection{Halo dimers and universality}
\label{ssec:universal}

\begin{figure}
\includegraphics[width=5.6in]{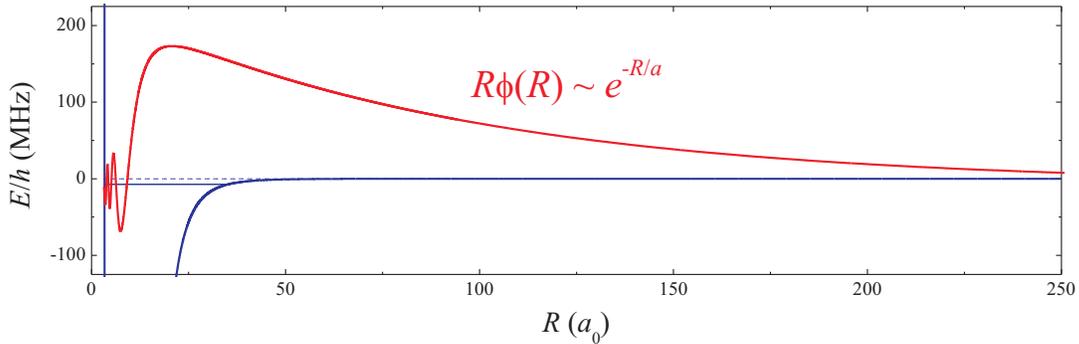}
\caption{Radial wavefunction of a halo dimer, where $R$ is the
interatomic distance, extending far into the classically forbidden
range where the wavefunction is proportional to $e^{-R/a}$. The
wavefunction is calculated for a Lennard-Jones potential and
$m_r=3$~a.m.u. \cite{LeRoy2007}. The potential depth is tuned such
that the highest vibrational level, in this case $v=5$, is in the
halo regime.}\label{fig:halowavefunction}
\end{figure}

For large scattering lengths $a$, the Feshbach molecule enters the
halo regime, where its binding energy is given by Eq.~\ref{Eb}. Its
magnetic field dependence is illustrated in the inset of
Fig.~\ref{fig:intro}. The halo state can be described in terms of a
single effective molecular potential having scattering length $a$.
In Fig.~\ref{fig:halowavefunction} a typical halo wavefunction is
shown as function of the interatomic distance $R$. This highlights
its remarkable property to extend far into the classically forbidden
range, which is at the heart of its universal properties. The
asymptotic form of the halo wavefunction is proportional to
$e^{-R/a}$ and the average distance between the two atoms is $a/2$;
see also Sec.~V of Chapter 11.

The halo regime requires $a$ to be much larger than the range of the
two-body potential. For ground state alkali atoms a characteristic
range can be described by the van der Waals length $R_{\rm vdw} =
\frac{1}{2}(2m_{\rm r}C_6/\hbar^2)^{1/4}$, where $C_6$ is the van
der Waals dispersion coefficient; see Chapter 6 and 10. The
corresponding van der Waals energy is $E_{\rm vdw}= \hbar^2/(2m_{\rm
r} R_{\rm vdw}^2)$. Thus a halo dimer can be defined through the
criteria $a\gg R_{\rm vdw}$ and $E_{\rm b}\ll E_{\rm vdw}$. For some
examples the van der Waals scales are given in
Table~\ref{table:RvdWEvdW}, where $R_{\rm vdw}$ is expressed in
units of the Bohr radius $a_0=0.529\times10^{-10}$~m.

The halo regime corresponds to the \emph{universal regime} for
$a>0$. The concept of universality is that the properties of any
physical system in which $|a|$ is much larger than the range of the
interaction is determined by $a$, and therefore all these systems
exhibit the same \emph{universal} behavior \cite{Braaten2006uif}. In
the universal regime the details of the short range interaction
become irrelevant because of the dominant long-range nature of the
wavefunction. The existence of the halo dimer at large positive $a$
is a manifestation of universality in two-body physics
\cite{Jensen2004sar}. Halo states are known in nuclear physics, with
the deuteron being a prominent example. In molecular physics, the He
dimer has for many years served as the prime example of a halo
state. Feshbach molecules with halo character are special to other
halo states in the sense that their properties are magnetically
tunable. In particular, one can scan from the universal regime of
halo dimers to the non-universal regime of non-halo Feshbach
molecules.

\begin{table}
\caption{Characteristic van der Waals scales $R_{\rm vdw}$ and
$E_{\rm vdw}$ for some selected examples.\label{table:RvdWEvdW}}
\begin{ruledtabular}
\begin{tabular}{c c c c c c}
                      & $^{6}$Li$_2$ & $^{40}$K$_2$ & $^{40}$K$^{87}$Rb & $^{85}$Rb$_2$ & $^{133}$Cs$_2$ \\
  \hline
$R_{\rm vdw}$~($a_0$) & 31           & 65           & 72                & 82            & 101              \\
$E_{\rm vdw}/h$~(MHz) & 614          & 21           & 13                & 6.2           & 2.7 \\
\end{tabular}
\end{ruledtabular}
\end{table}

In principle, every Feshbach resonance features a halo region. In
practice, however, only a few resonances are suited to
experimentally realize this interesting regime. Those are broad
resonances with large widths $|\Delta|$ (typically $\gg 1$\,G) and
large background scattering lengths $|a_{\rm bg}|$ that exceed
$R_{\rm vdw}$ \cite{Kohler2006poc,Chin2008fri}. Prominent examples
are found in the two fermionic systems $^6$Li and $^{40}$K, and the
bosonic systems $^{39}$K, $^{85}$Rb and $^{133}$Cs, some of which
are discussed in Sec.~IV of Chapter 6 and Sec.~V of Chapter 11.

Another universal feature of a halo dimer regards spontaneous
dissociation, which is possible when its atomic constituents are not
in the lowest internal states and open decay channels are present.
In the halo regime the dissociation rate scales as $a^{-3}$, as has
been observed for $^{85}$Rb$_2$ \cite{Thompson2005sdo}, and can be
understood as a direct consequence of the large halo wavefunction
\cite{Kohler2005sdo}.

\subsection{Collisional properties and few-body physics}
\label{ssec:halocoll}

The concept of universality extends beyond two-body physics to
few-body phenomena, including the low-energy scattering properties
of halo dimers. The \emph{inelastic} collisional properties are
important as inelastic collisions usually lead to trap loss,
determining the lifetime of a sample of trapped halo dimers. At the
same they provide a convenient experimental observable to study
few-body physics. In general, the loss of dimers can be described by
the following rate equation:
\begin{equation}
\dot{n}_{\rm D} = -\alpha n^2_{\rm D}-\beta n_{\rm D} n_{\rm A},
\end{equation}
where $\alpha$ and $\beta$ are the loss rate coefficients for
dimer-dimer and atom-dimer collisions, respectively, and $n_{\rm A}
(n_{\rm D})$ is the atomic (molecular) density. Fast trap loss has
been observed for Feshbach molecules in the non-universal regime,
with $\alpha$ and $\beta$ on the order of $10^{-11}$-$10^{-10}$
cm$^3$s$^{-1}$; see Sec.~III of Chapter 3. The loss coefficient
$\alpha$ is most conveniently measured using a pure molecular
sample. To determine $\beta$ an atom-dimer mixture is required, the
atom number greatly exceeding the molecule number being beneficial.

In the universal regime, simple scaling laws for $\alpha$ and
$\beta$ can be derived. For halo dimers composed of two identical
bosons $\beta$ scales linearly with $a$
\cite{Braaten2004edr,DIncao2005sls}, resulting in an enhancement of
dimer loss near a Feshbach resonance. In contrast, for halo dimers
made of two fermions in different spin states an $a^{-2.55}$ and an
$a^{-3.33}$ scaling are derived for $\alpha$ and $\beta$
\cite{Petrov2004wbm}, respectively, connected to the collisional
stability of the molecular sample near a Feshbach resonance, as
discussed in Sec.~\ref{ssec:statistic}. Universal scaling laws are
also derived for other atom-dimer systems
\cite{Zirbel2008cso,DIncao2008som}, as well as for the related
problem of three-body recombination in atomic samples
\cite{DIncao2005sls,DIncao2008som}.

\begin{figure}
\includegraphics[width=3.2in]{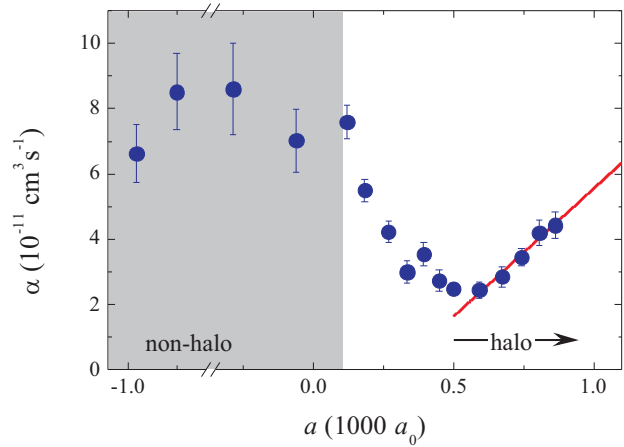}
\caption{Experimental study of the interaction between halo dimers.
The relaxation rate coefficient $\alpha$ for inelastic dimer-dimer
scattering is measured as function of the scattering length $a$ for
a pure sample of optically trapped Cs Feshbach molecules. The
experiment makes use of the wide tunability of the scattering length
for Cs, in particular into the halo regime. The solid line is a
linear fit to the data in the region $a\geq500\,a_0$. The shaded
region indicates the $a<R_{\rm vdw}$ regime, where the Feshbach
molecules are not in a halo state. Adapted from
Ref.~\cite{Ferlaino2008cbt}.} \label{fig:Dimerdimerscattering}
\end{figure}

Inelastic dimer-dimer scattering represents an elementary four-body
process. For four identical bosons a universal prediction has so far
not been derived, as the four-body problem is substantially more
challenging than the three-body problem. Experimentally one can
investigate this process by measuring $\alpha$ in a pure molecular
sample of halo dimers made of identical bosons, which has been done
in the case of Cs$_2$ \cite{Ferlaino2008cbt}. The results are given
in Fig.~\ref{fig:Dimerdimerscattering}, showing a strong scattering
length dependence of $\alpha$. A pronounced loss minimum around
$500a_0$ is observed, followed by a linear increase towards larger
$a$. The interpretation of this striking behavior is still an open
issue and might stimulate progress in the theoretical description of
the four-body problem.

For the \emph{elastic} scattering properties, universal scaling laws
have been derived for fermion-composed halo dimers. Both the
atom-dimer and dimer-dimer scattering lengths simply scale with the
atom-atom scattering length $a$, namely as $1.2\,a$
\cite{Skorniakov1957tbp} and $0.6\,a$ \cite{Petrov2004wbm},
respectively. Therefore the elastic cross sections, being
proportional to the square of these scattering lengths, are very
large in the halo region. This leads to fast thermalization in an
atom-dimer mixture and a pure dimer sample, opening the possibility
of evaporative cooling. For boson-composed halo dimers the few-body
problem is more complicated and universal predictions for their
elastic scattering properties are so far lacking.

\subsection{Efimov three-body states}
\label{ssec:efimov}

Efimov quantum states in a system of three identical bosons
\cite{Efimov1970ela,Efimov1971wbs} are a paradigm for universal
few-body physics. These states have attracted considerable interest,
fueled by their bizarre and counter-intuitive properties and by the
fact that they had been elusive to experimentalists for more than 35
years. In 2006, experimental evidence for Efimov states in an
ultracold gas of Cs atoms was reported \cite{Kraemer2006efe}. In the
context of ultracold quantum gases, Efimov physics manifests itself
in three-body decay properties, such as resonances in the three-body
recombination and atom-dimer relaxation loss rates. These resonances
appear on top of the non-resonant ``background" scattering behavior,
given by the universal scaling laws discussed in
Sec.~\ref{ssec:halocoll}.

\begin{figure}
\includegraphics[width=3.4in]{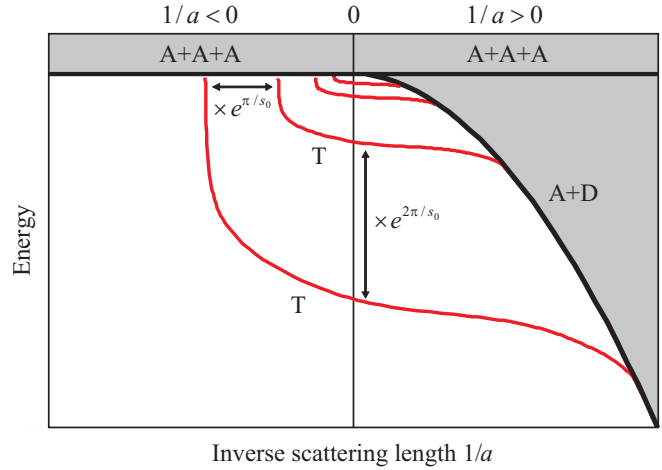}
\caption{Efimov's scenario. Appearance of an infinite series of
weakly bound Efimov trimer states (T) for resonant two-body
interaction, showing a logarithmic-periodic pattern with universal
scaling factors $e^{\pi/s_0}$ and $e^{2\pi/s_0}$ for the scattering
length and the binding energy, respectively. The binding energy is
plotted as a function of the inverse two-body scattering length
$1/a$. The gray regions indicate the scattering continuum for three
atoms (A+A+A) and for an atom and a halo dimer (A+D). To illustrate
the series of Efimov states, the universal scaling factor is
artificially set to 2.} \label{fig:EfimovScenario}
\end{figure}

Efimov's scenario is illustrated in Fig.~\ref{fig:EfimovScenario},
showing the energy spectrum of the three-body system as a function
of the inverse scattering length $1/a$. For $a < 0$, the natural
three-body dissociation threshold is at zero energy. States below
are trimer states and states above are continuum states of three
free atoms. For $a > 0$, the dissociation threshold is given by the
binding energy of the halo dimer; at this threshold a trimer
dissociates into a halo dimer and an atom. Efimov predicted an
infinite series of weakly bound trimer states with universal scaling
behavior. When the scattering length is increased by a universal
scaling factor $e^{\pi/s_0}$, a new Efimov state appears, which is
just larger by this factor and has a weaker binding energy by a
factor $e^{2\pi/s_0}$. For three identical bosons
$s_0\approx1.00624$, so that the scaling factors for the scattering
length and the binding energy are 22.7 and $22.7^2\approx 515$,
respectively. For $a < 0$, Efimov states are ``Borromean" states
\cite{Jensen2004sar}, which means that a weakly bound three-body
state exists in the absence of a weakly bound two-body state. This
property that three quantum objects stay together without pairwise
binding is part of the bizarre nature of Efimov states.

\begin{figure}
\includegraphics[width=3.2in]{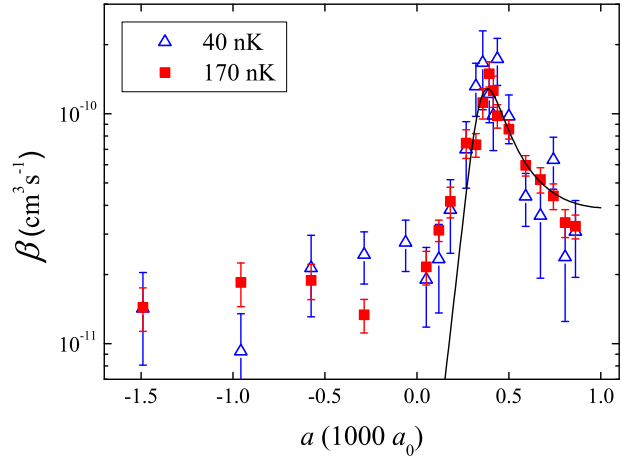}
\caption{Observation of a scattering resonance in an ultracold,
optically trapped mixture of Cs atoms and Cs$_2$ halo dimers. As for
Fig.~\ref{fig:Dimerdimerscattering}, one benefits from the wide
tunability of the scattering length provided by Cs. The data show
the loss rate coefficient for atom-dimer relaxation $\beta$ for
temperatures of 40(10)~nK and 170(20)~nK. A pronounced resonance is
observed at about +400\,$a_0$, corresponding to a magnetic field of
25\,G. The solid curve is a fit of an analytic model from effective
field theory \cite{Braaten2006uif} to the data for $a>R_{\rm vdw}$.
Adapted from Ref.~\cite{Knoop2008ooa}.}
\label{fig:atomdimerresonance}
\end{figure}

The Efimov trimers influence the three-body scattering properties.
When an Efimov state intersects the continuum threshold for $a < 0$
three-body recombination loss is enhanced
\cite{Esry1999rot,Braaten2001tbr}, as the resonant coupling of three
atoms to an Efimov state opens up fast decay channels into deeply
bound dimer states plus a free atom. Such an Efimov resonance has
been observed in an ultracold, thermal gas of Cs atoms
\cite{Kraemer2006efe}. For $a>0$ a similar phenomenon is predicted,
namely an atom-dimer scattering resonance at the location at which
an Efimov state intersects the atom-dimer threshold
\cite{Nielsen2002eri,Braaten2007rdr}. Resonance enhancement of
$\beta$ has been observed in a mixture of Cs atoms and Cs$_2$ halo
dimers \cite{Knoop2008ooa}; see Fig.~\ref{fig:atomdimerresonance}.
The asymmetric shape of the resonance can be explained by the
background scattering behavior, which here is a linear increase as
function of $a$.

Efimov physics impacts not only the scattering properties of three
identical bosons, but any three-body system in which at least two of
the three pair-wise two-body interactions are large. Interestingly,
depending on the mass ratio of the different components and the
particle statistics, the scaling factor $e^{\pi/s_0}$ can be much
smaller than 22.7, facilitating the observation of successive Efimov
resonances in future experiments \cite{DIncao2006eto}.

\subsection{Molecular BEC}
\label{ssec:fermi}

The achievement of Bose-Einstein condensations of molecules in out
of degenerate atomic Fermi gases is arguably one of the most
remarkable results in the field of ultracold quantum gases. The
experimental realization was demonstrated in ultracold Fermi gases
of $^6$Li and $^{40}$K and turned out to be an excellent starting
point to investigate the so-called BEC-BCS crossover and the
properties of strongly interacting Fermi gases
\cite{Varenna2006,Giorgini2007tou}. The realization of an mBEC of
Feshbach molecules is only possible in the halo regime, where the
collisional properties are favorable; see Sec.~\ref{ssec:halocoll}.

In a spin mixture of $^6$Li, containing the lowest two hyperfine
states, the route to mBEC is particularly simple
\cite{Jochim2003bec}. Evaporative cooling towards BEC can be
performed in an optical dipole trap at a constant magnetic field in
the halo region near the Feshbach resonance. In the initial stage of
evaporative cooling the gas is purely atomic and molecules are
produced via three-body recombination; see Sec.~\ref{ssec:assoc}.
With decreasing temperature the atom-molecule equilibrium favors the
formation of molecules and a purely molecular sample is cooled down
to BEC. The large atom-dimer and dimer-dimer scattering lengths
along with strongly suppressed relaxation loss facilitate an
efficient evaporation process. In this way, mBECs are achieved with
a condensate fraction exceeding 90\%.

\begin{figure}
\includegraphics[width=3.5in]{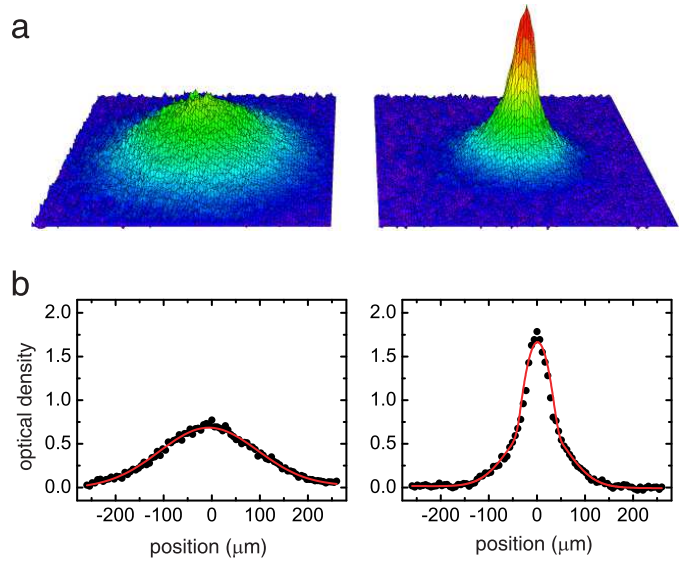}
\caption{Emergence of a molecular BEC in an ultracold Fermi gas of
$^{40}$K atoms, observed in time-of-flight absorption images. The
density distribution on the left-hand side (upper graph, 2D surface
plot; lower graph, 1D cross section) was taken for a Fermi gas which
was cooled down to 19\% of the Fermi temperature. After ramping
across the Feshbach resonance no mBEC was observed as the sample was
too hot. The density distribution on the right-hand side was
observed for a colder sample at 6\% of the Fermi temperature. Here
the ramp across the Feshbach resonance resulted in a bimodal
distribution, revealing the presence of an mBEC with a condensate
fraction of 12\%. Reprinted by permission from Macmillan Publishers
Ltd: Nature (London), Greiner \emph{et al.} \cite{Greiner2003eoa},
copyright 2003.} \label{fig:mbec_jila}
\end{figure}

The experiments in $^{40}$K followed a different approach to achieve
mBEC \cite{Greiner2003eoa}. For $^{40}$K the halo dimers are less
stable because of less favorable short-range three-body interaction
properties. Therefore the sample is first cooled above the Feshbach
resonance, where $a$ is large and negative, to achieve a deeply
degenerate atomic Fermi gas. A sweep across the Feshbach resonance
then converts the sample into a partially condensed cloud of
molecules. The emergence of the mBEC in $^{40}$K is shown in
Fig.~\ref{fig:mbec_jila}.

For very large $a$ the size of the halo dimers becomes comparable to
the interparticle spacing, and the properties of the fermion pairs
start to be determined by many-body physics. For $a<0$ two-body
physics no longer supports the weakly bound molecular state and
pairing is entirely a many-body effect. In particular, in the limit
of weak interactions on the $a<0$ side of the resonance, pairing can
be understood in the framework of the well-established
Bardeen-Cooper-Schrieffer (BCS) theory, developed in the 1950s to
describe superconductivity. Here the fermionic pairs are called
Cooper pairs. The BEC and BCS limits are smoothly connected by a
crossover regime where the gas is strongly interacting. This BEC-BCS
crossover has attracted considerable attention in many-body quantum
physics \cite{Varenna2006,Giorgini2007tou,Bloch2008mbp}. A
theoretical description of this challenging problem is very
difficult and various approaches have been developed. With tunable
Fermi gases, a unique testing ground has become available to
quantitatively investigate the crossover problem.

\section{Toward ground-state molecules}
\label{sec:towardground}
\newcommand{\Rb}{$^{87}$Rb$_2 $ }
\newcommand{\Cs}{$^{133}$Cs$_2 $ }
\newcommand{\fb}{Fermi-Bose mixture }
\newcommand{\KRb}{$^{40}$K$^{87}$Rb }
\begin{figure}
\includegraphics[width=6.0in]{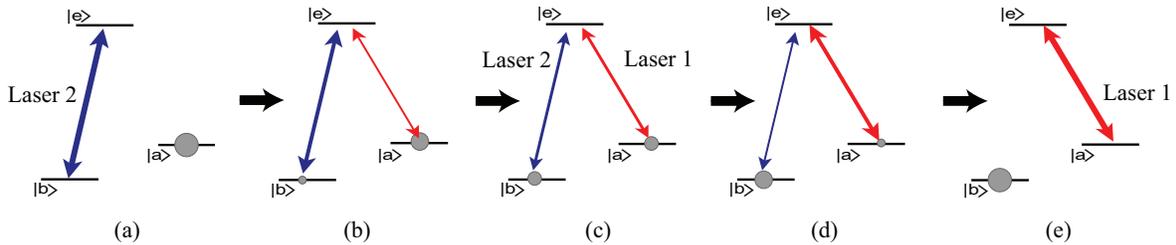}
\caption{Illustration of the basic idea of STIRAP. Transfer from state
$|a\rangle$ to state $|b\rangle$, keeping the population always in a dark state
(see Eq.\,\ref{eq:DS}). Initially $\Omega_{2}=1$ and $\Omega_{1}=0$, and finally
$\Omega_{2}=0$ and $\Omega_{1}=1$.} \label{fig:3level}
\end{figure}

Ultracold gases of ground-state molecules hold great prospects for novel quantum systems with more complex interactions than existing in atomic quantum gases. For the experimental preparation of such molecular systems, the collisional stability is an important prerequisite. This makes rovibrational ground states the prime candidates, as vibrational relaxation is energetically suppressed.

Stable BECs of dimers may for instance serve as a starting point to synthesize mBECs of more complex constituents. A particular motivation arises from the large electric dipole moments of heteronuclear dimers; see Chapter 2. This leads to the long-range dipole-dipole interaction, which is anisotropic and can be modified by an electric field. As a result, a rich variety of phenomena can expected, with conceptual challenges and intriguing experimental opportunities. Proposals range from the study of new quantum phases and  quantum simulations of condensed-matter system (Chapter 12), to fundamental physics tests (Chapters~15 and 16) and schemes for quantum information processing (Chapter 17).

 The method of STImulated Raman Adiabatic Passage (STIRAP) has attracted considerable interest as a highly efficient way to coherently transfer atom pairs or Feshbach molecules into more deeply bound state \cite{Jaksch2002coa}. The method offers a high transfer efficiency without heating of the molecular sample, thus allowing to preserve the high phase-space density of an ultracold gas. Several groups are actively pursuing such experiments with the motivation to create quantum gases of collisionally stable ground-state molecules.

\subsection{Stimulated Raman adiabatic passage}
\label{ssec:stirap}

The basic idea of STIRAP to transfer population between two quantum states relies on a particularly clever implementation of a coherent two-photon Raman transition, involving a dark state during the transfer.
For a detailed description of its principles and an overview of early applications, the reader may refer to Ref.~\cite{Bergmann1998cpt}.

Let us consider a three-level system with $|a\rangle$ and $|b\rangle$ representing two different ground state levels, and $|e\rangle$ being an electronically excited state, as shown in Fig.\,\ref{fig:3level}.
Laser~1~(2) couples the state $|a\rangle$ ($|b\rangle$) to the state
$|e\rangle$ and the Rabi frequency $\Omega_{1}$  ($\Omega_{2} $)
describes the corresponding coupling strength \cite{Kuklinski1989apt}. The Rabi frequency is defined as $\mathbf{d} \cdot \mathbf{E}/\hbar$, with $\mathbf{E}$ being the electric-field amplitude of the laser field and $\mathbf{d}$ the dipole matrix element. The states $|a\rangle$ and $|b\rangle$ are long lived, whereas the excited state $|e\rangle$ can undergo spontaneous decay into lower states.

An essential property of such a coherently coupled three-level system
is the existence of a dark state $|D\rangle$ as an eigenstate of the system.  This state generally occurs if both laser fields have the same resonance detuning with respect to the corresponding transition, i.e.\ if the two-photon detuning is zero. The state is \emph{dark} in the sense that it is decoupled from the excited state $|e\rangle$ and thus not influenced by its radiative decay. The dark state can be understood as a coherent superposition of state $|a\rangle$ and state $|b\rangle$,
\begin{equation}
  \label{eq:DS}
  |D\rangle = \frac{1}{\sqrt{{\Omega_1}^2 + {\Omega_2}^2}}
\left(\Omega_2 |a\rangle - \Omega_1 |g\rangle\right) ,
\end{equation}
with state $|e\rangle$ being not involved.

The key idea of STIRAP is to slowly change $\Omega_1$ and $\Omega_2$ in a way that the system is always kept in a dark state. This avoids losses caused by spontaneous light scattering in the excited state $|e\rangle$ \cite{Oreg1984afi}. One may intuitively expect that the transfer occurs by first applying Laser~1 and then Laser~2, as in a conventional two-photon Raman transition. STIRAP instead uses a counterintuitive laser pulse scheme, in which Laser~2 is applied first.
Initially, only state $|a\rangle$ is populated (Fig.\,\ref{fig:3level}a); this corresponds to the dark state $|D\rangle$ if only Laser~2 is active. If one now slowly introduces Laser~1, the dark state $|D\rangle$ begins to evolve into a superposition of $|a\rangle$ and $|b\rangle$ (Fig.\,\ref{fig:3level}b). By adiabatically ramping down the power of Laser~2 and ramping up the intensity of Laser~1, the character of the superposition state changes and it acquires an increasing $|b\rangle$ admixture (Fig.\,\ref{fig:3level}c-d).
When finally Laser~1 is on and Laser~2 is turned off, the population is completely transfered to the state $|b\rangle$  (Fig.\,\ref{fig:3level}e). This adiabatic passage through the dark state $|D\rangle$ does not involve any population in the radiatively decaying state $|e\rangle$. We point out that the relative phase coherence of the two laser fields is an essential requirement for STIRAP. Moreover, it is important to note that the coherent sequence can be time-reversed to achieve transfer from $|b\rangle$ to $|a\rangle$.

\begin{figure}
\includegraphics[width=3.5in]{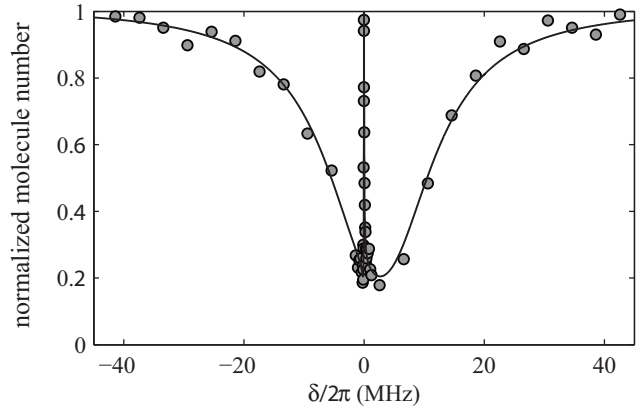}
\vspace{-6mm}
\caption{Dark resonance observed as a signature of a dark state between two different vibrational levels in a trapped sample of $^{87}$Rb$_2$ Feshbach molecules. Here both laser fields are turned on simultaneously, and the frequency of Laser~1 is varied while Laser~2 is kept at a fixed frequency, thus varying the two-photon detuning $\delta$. The population of the optically excited state $|e\rangle$ leads to a strong background loss caused by spontaneous decay into other levels in a region corresponding to the single-photon transition linewidth. If the two-photon resonance condition is met ($\delta =0$), a suppression of loss occurs in a very narrow frequency range, which indicates the decoupling from the excited state. The behavior can be modeled on the basis of a three-state system, which provides full information on the relevant experimental parameters. In practice, the study of such a dark resonance is an important step to implement STIRAP.
From Ref.~\cite{Winkler2007cot}.} \label{fig:dark}
\end{figure}

\subsection{STIRAP experiments}
\label{ssec:stirapexpts}

In the experiments to create quantum gases of collisionally stable ground-state molecules, the initially populated state $|a\rangle$ corresponds to a Feshbach molecule and $|b\rangle$ is a deeply bound molecular state. The state $|e\rangle$ is a carefully chosen electronically excited level, which has to fulfil two important conditions. To obtain large enough optical Rabi frequencies $\Omega_1$ and $\Omega_2$ the state must provide sufficiently strong coupling to both states $|a\rangle$ and $|b\rangle$, demanding sufficiently strong dipole matrix elements besides the technical requirement of sufficiently intense laser fields at the particular wavelengths. Moreover, state $|e\rangle$ needs to be well separated from other excited states, as a significant coupling to these would deteriorate the dark character of the superposition state.

Proof-of-principle experiments have been performed on homonuclear $^{87}$Rb$_2$ molecules \cite{Winkler2007cot} and heteronuclear  \KRb  molecules \cite{Ospelkaus2008est}, demonstrating the transfer by one or a few vibrational quanta. The STIRAP transfer of $^{87}$Rb$_2$ started with trapped Feshbach molecules. In order to characterize and optimize the experimental parameters, the study of a ``dark resonance'' was an essential step; see Fig.~\ref{fig:dark}. STIRAP could then be efficiently implemented to transfer the molecules from the last bound vibrational level ($E_{\rm b}= h \times 24$~MHz) to the second-to-last state ($h \times 637$~MHz) with an efficiency close to 90$\%$. In the experiment with the heteronuclear \KRb molecules, STIRAP was demonstrated with final molecular binding energies of up to $h \times 10$~GHz.

Subsequent experiments then moved on to more deeply bound molecules and in particular to the rovibrational ground states of the triplet and the singlet potential. To bridge a large energy range with STIRAP the experiments become more challenging both for technical and for physical reasons. In particular, the optical transition matrix elements become an important issue, as they enter the optical Rabi frequencies. In molecular physics, the Franck-Condon principle states that an optical transition does not change the internuclear separation. As a consequence, the overlap of the wavefunctions between ground state and excited state, quantified by the so-called Franck-Condon factors, crucially enters into the matrix elements. A general rule of thumb can be given for finding a large Franck-Condon overlap based on a simple classical argument. The dominant part of the wavefunction occurs near the turning points of the classical oscillatory motion in a specific molecular potential. A large Franck-Condon factor for an optical transition can be expected if the classical turning points approximately coincide for the ground and excited state.

\begin{figure}
\includegraphics[width=6in]{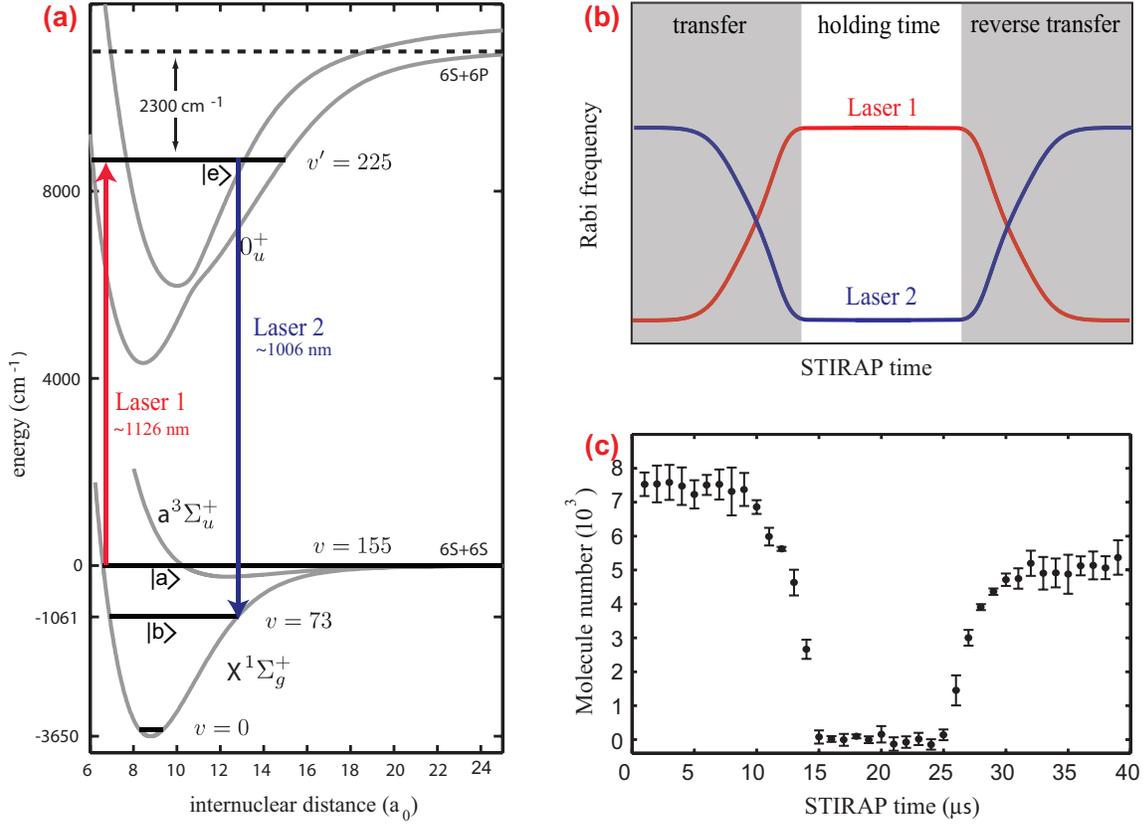}
\caption{STIRAP experiment starting with ultracold \Cs Feshbach molecules. In (a) the relevant molecular potential curves and energy levels are schematically shown. The vertical arrows show the optical transitions and their positions indicate the internuclear distances where maximum Franck-Condon overlap is expected near the classical turning points. The variation of the Rabi frequencies during
the pulse sequence is illustrated in (b). The experimental results in (c) show the measured molecule population in $|a\rangle$ as a function of time through a double-STIRAP sequence. Adapted from Ref.~\cite{Danzl2008qgo}.} \label{fig:gs}
\end{figure}

Molecules with large binding energies of about $h\times 32$~THz were demonstrated in an experiment on \Cs \cite{Danzl2008qgo}. The experiment starts with the creation of a pure sample of Feshbach molecules in a weakly bound state $|a\rangle$ with a binding energy of $h \times 5\,$MHz. Here the standard methods for association and purification are applied as described in Sec.~\ref{sec:makedetect}. A suitable Franck-Condon overlap was experimentally found for the particular choice of $|e\rangle$ that is depicted in  Fig.\,\ref{fig:gs}(a). The vertical arrows indicate the optical transitions at internuclear separations where maximum wavefunction overlap occurs. The applied timing sequence is shown in Fig.\,\ref{fig:gs}(b). In a first STIRAP pulse sequence, the molecules are transferred from $|a\rangle$ to $|b\rangle$ within 15\,$\mu$s. After a variable holding time in state $|b\rangle$, the deeply bound molecules are reconverted into Feshbach molecules by a time-reversed STIRAP pulse sequence. The Feshbach molecules are then detected by the usual dissociative imaging.  The experimental results in Fig.\,\ref{fig:gs}(c) show that 65\% of the initial number of molecules reappear after the full double-STIRAP sequence, pointing to a single STIRAP efficiency of about 80\%.

In the final stage of preparation of the present manuscript, breakthroughs have been achieved on the creation of rovibrational ground-state molecules in three different experiments. While Ref.\,\cite{Ni2008ahp} reports on polar \KRb molecules in both the ground state of the triplet potential ($E_{\rm b} = h \times 7.2$\,THz) and the one of the singlet potential ($h \times 125$\,THz), Ref.\,\cite{Lang2008umi} demonstrates homonuclear \Rb molecules in the rovibrational ground state of the triplet potential ($h \times 7.0$\,THz). A further experiment \cite{Naegerl2008} has succeeded in the production of \Cs molecules in the rovibrational ground state of the singlet potential ($h \times 109$\,THz).

The successful \KRb experiment also provides a further remarkable demonstration of the importance of the Franck-Condon overlap arguments. At first glance, it may be surprising that the huge binding energy difference to the singlet ground state could be bridged in a single STIRAP step. However, an excited molecular state $|e\rangle$ could be found for which the outer classical turning point provides good overlap with the Feshbach molecular state, while the internuclear distance at the inner turning point matches the rovibrational ground state. In the absence of such a fortunate coincidence, transfer may not be feasible in a single STIRAP sequence. Two consecutive STIRAP steps involving different excited states \cite{Jaksch2002coa, Naegerl2008} promise a general way to achieve efficient ground-state transfer for any homo- or heteronuclear Feshbach molecule.

\section{Further developments and concluding remarks}
\label{sec:conclude}

In the rapidly developing field of cold molecules, Feshbach molecules play a particular role as the link to research on ultracold quantum matter. In this Chapter, we have presented the basic experimental techniques and discussed some major developments in the field. In this final Section, we briefly point to some other recent developments not discussed so far, adding further interesting aspects to our overview of the research field.

Intriguing connections to condensed-matter physics can be found when Feshbach molecules are trapped in optical lattices. In this spatially periodic environment, a Feshbach molecule can be used both as a well-controllable source of correlated atom pairs, and as an efficient tool to detect such pairs. A recent experiment \cite{Winkler2006rba} shows how Feshbach molecules, prepared in a three-dimensional lattice, are converted into ``repulsively bound pairs'' of atoms when forcing their dissociation through a Feshbach ramp. Such atom pairs stay together and jointly hop between different sites of the lattice, because the atoms repel each other. This counterintuitive behavior is due to the fact that the band gap of the lattice does not provide any available states for taking up the interaction energy.

\begin{figure}
\includegraphics[width=3in]{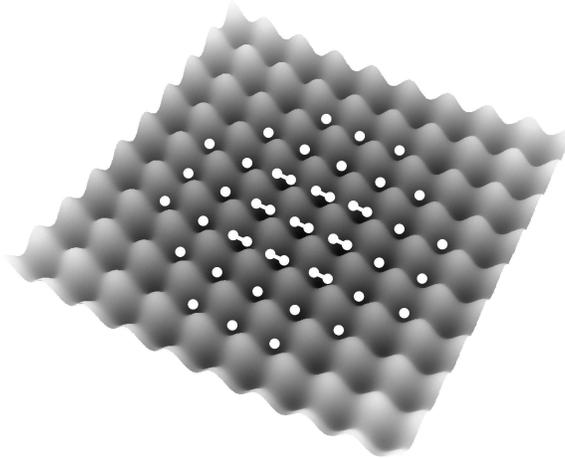}
\caption{Illustration of a quantum state with one molecule at each site of an optical lattice. In the central region of the three-dimensional lattice one finds exactly one molecule per site, trapped in the vibrational ground state. Such a state was prepared with $^{87}$Rb$_2$ Feshbach molecules. Adapted by permission from Macmillan Publishers Ltd: Nature Phys.\ (London), Volz {\it et al.} \cite{Volz2006pqs}, copyright 2006.} \label{fig:mmott}
\end{figure}

Another fascinating example along the lines of condensed-matter physics is a quantum state with exactly one Feshbach molecule per lattice site, as illustrated in Fig.~\ref{fig:mmott}. Such a state has vanishing entropy and closely resembles a Mott-insulator state \cite{Greiner2002qpt}. It offers an excellent starting point for experiments on strongly correlated many-body states (Chapter 12) and on quantum information processing (Chapter 17), and for the production of a BEC of molecules in the internal ground state \cite{Jaksch2002coa}. In the latter case, the basic idea is to convert the collisionally unstable Feshbach molecules into collisionally stable molecules in the rovibrational ground state. Then the lattice is no longer needed for isolating the molecules from each other and dynamical melting of the ordered state by ramping down the lattice may eventually lead to mBEC.

Novel phenomena can also be observed with Feshbach molecules in low-dimensional trapping schemes, such as one-dimensional tubes realized in two-dimensional optical lattices. Here a two-body bound state can exist also for negative scattering lengths, contrary to the situation in free space \cite{Moritz2005cim}.
Low-dimensional traps also offer very interesting environments to create strongly correlated many-body regimes. Counterintuitively, the sensitivity of Feshbach molecules to inelastic decay can inhibit loss and lead to a stable correlated state, as recently demonstrated with $^{87}$Rb$_2$ in a one-dimensional trap \cite{Syassen2008sdi}.

Various mixtures of ultracold atomic species offer a wide playground for ultracold heteronuclear molecules, and the basic interaction properties of many different combinations involving bosonic and fermionic atoms are currently being explored. A new twist to the field is given by ultracold Fermi-Fermi systems, which have recently been realized in mixtures of $^6$Li and $^{40}$K \cite{Taglieber2008qdt, Wille2008eau}. Molecule formation is such systems will lead to bosonic molecules, which in the many-body context may lead to new types of pair-correlated states and novel types of strongly interacting fermionic superfluids.

Another interesting question is whether more complex ultracold molecules can be created than simple dimers. A first step into this direction is the observation of scattering resonances between ultracold dimers \cite{Chin2005oof} which have been found in the collisional decay of an optically trapped sample of $^{133}$Cs$_2$ molecules and interpreted as a result of a resonant coupling to tetramer states.

All the examples presented in this Chapter highlight how great progress is currently being made to fully control the internal and external degrees of freedom of various types of ultracold molecules under conditions near quantum degeneracy. This will soon enable many new applications, ranging from high-precision measurements and quantum computation to the exploration of few-body physics and novel correlated many-body quantum states. Several major research themes are obvious and will undoubtedly lead to great success, but we are also confident that the field also holds the potential for many surprises and developments which we can hardly imagine now.

\section*{Acknowledgments}

We would like to thank all members of the Innsbruck group ``Ultracold Atoms and Quantum Gases'' (\verb"www.ultracold.at") for contributing to our research program on Feshbach molecules. In particular, we would like to thank Johannes Hecker Denschlag and Hanns-Christoph N\"agerl for the long-standing collaboration on this topic. We are also indebted to Cheng Chin, Paul Julienne, and Eite Tiesinga for the many insights gained when jointly preparing a review article on Feshbach resonances together with one of us (R.G.). We are grateful to the Austrian Science Fund (FWF) for supporting our various projects related to ultracold molecules. F.F.\ is a Lise-Meitner fellow of the FWF, and S.K.\ is supported by the European Commission with a Marie Curie Intra-European Fellowship.


\begin{thebibliography}{10}

\expandafter\ifx\csname natexlab\endcsname\relax\def\natexlab#1{#1}\fi
\expandafter\ifx\csname bibnamefont\endcsname\relax
  \def\bibnamefont#1{#1}\fi
\expandafter\ifx\csname bibfnamefont\endcsname\relax
  \def\bibfnamefont#1{#1}\fi
\expandafter\ifx\csname citenamefont\endcsname\relax
  \def\citenamefont#1{#1}\fi
\expandafter\ifx\csname url\endcsname\relax
  \def\url#1{\texttt{#1}}\fi
\expandafter\ifx\csname urlprefix\endcsname\relax\def\urlprefix{URL }\fi
\providecommand{\bibinfo}[2]{#2}
\providecommand{\eprint}[2][]{\url{#2}}

\bibitem[{\citenamefont{Chu}(1998)}]{Chu1998nlt}
\bibinfo{author}{\bibfnamefont{S.}~\bibnamefont{Chu}}, \bibinfo{title}{``Nobel Lecture: The manipulation of neutral particles"}, \bibinfo{journal}{Rev.
  Mod. Phys.} \textbf{\bibinfo{volume}{70}}, \bibinfo{pages}{685}
  (\bibinfo{year}{1998}).

\bibitem[{\citenamefont{Cohen-Tannoudji}(1998)}]{Cohentannoudji1998nlm}
\bibinfo{author}{\bibfnamefont{C.~N.} \bibnamefont{Cohen-Tannoudji}}, \bibinfo{title}{``Nobel Lecture: Manipulating atoms with photons"}, \bibinfo{journal}{Rev. Mod. Phys.} \textbf{\bibinfo{volume}{70}},
  \bibinfo{pages}{707} (\bibinfo{year}{1998}).

\bibitem[{\citenamefont{Phillips}(1998)}]{Phillips1998nll}
\bibinfo{author}{\bibfnamefont{W.~D.} \bibnamefont{Phillips}}, \bibinfo{title}{``Nobel Lecture: Laser cooling and trapping of neutral atoms"}, \bibinfo{journal}{Rev. Mod. Phys.} \textbf{\bibinfo{volume}{70}},
  \bibinfo{pages}{721} (\bibinfo{year}{1998}).

\bibitem[{\citenamefont{Cornell and Wieman}(2002)}]{Cornell2002nlb}
\bibinfo{author}{\bibfnamefont{E.~A.} \bibnamefont{Cornell}} \bibnamefont{and}
  \bibinfo{author}{\bibfnamefont{C.~E.} \bibnamefont{Wieman}}, \bibinfo{title}{``Nobel Lecture: Bose-Einstein condensation in a dilute gas, the first
    70 years and some recent experiments"}, \bibinfo{journal}{Rev. Mod. Phys.} \textbf{\bibinfo{volume}{74}},
  \bibinfo{pages}{875} (\bibinfo{year}{2002}).

\bibitem[{\citenamefont{Ketterle}(2002)}]{Ketterle2002nlw}
\bibinfo{author}{\bibfnamefont{W.}~\bibnamefont{Ketterle}}, \bibinfo{title}{``Nobel lecture: When atoms behave as waves: Bose-Einstein condensation
    and the atom laser"},  \bibinfo{journal}{Rev. Mod. Phys.} \textbf{\bibinfo{volume}{74}},
  \bibinfo{eid}{1131} (\bibinfo{year}{2002}).

\bibitem[{\citenamefont{Arimondo et~al.}(1992)\citenamefont{Arimondo, Phillips,
  and Strumia}}]{Varenna1991}
\bibinfo{editor}{\bibfnamefont{E.}~\bibnamefont{Arimondo}},
  \bibinfo{editor}{\bibfnamefont{W.~D.} \bibnamefont{Phillips}},
  \bibnamefont{and} \bibinfo{editor}{\bibfnamefont{F.}~\bibnamefont{Strumia}},
  eds., \emph{\bibinfo{title}{Laser Manipulation of Atoms and Ions}}
  (\bibinfo{publisher}{North Holland, Amsterdam}, \bibinfo{year}{1992}),
  \bibinfo{note}{{P}roceedings of the International School of Physics ``Enrico
  Fermi'', Course CXVIII, Varenna, 9-19 July 1991}.

\bibitem[{\citenamefont{Metcalf and van~der Straten}(1999)}]{Metcalf1999book}
\bibinfo{author}{\bibfnamefont{H.~J.} \bibnamefont{Metcalf}} \bibnamefont{and}
  \bibinfo{author}{\bibfnamefont{P.}~\bibnamefont{van~der Straten}},
  \emph{\bibinfo{title}{Laser Cooling and Trapping}}
  (\bibinfo{publisher}{Springer, New York}, \bibinfo{year}{1999}).

\bibitem[{\citenamefont{Bize et~al.}(2005)\citenamefont{Bize, Laurent, Abgrall,
  Marion, Maksimovic, Cacciapuoti, Gr{\"u}nert, Vian, Pereira~dos Santos,
  Rosenbusch et~al.}}]{Bize2005cac}
\bibinfo{author}{\bibfnamefont{S.}~\bibnamefont{Bize}},
  \bibinfo{author}{\bibfnamefont{P.}~\bibnamefont{Laurent}},
  \bibinfo{author}{\bibfnamefont{M.}~\bibnamefont{Abgrall}},
  \bibinfo{author}{\bibfnamefont{H.}~\bibnamefont{Marion}},
  \bibinfo{author}{\bibfnamefont{I.}~\bibnamefont{Maksimovic}},
  \bibinfo{author}{\bibfnamefont{L.}~\bibnamefont{Cacciapuoti}},
  \bibinfo{author}{\bibfnamefont{J.}~\bibnamefont{Gr{\"u}nert}},
  \bibinfo{author}{\bibfnamefont{C.}~\bibnamefont{Vian}},
  \bibinfo{author}{\bibfnamefont{F.}~\bibnamefont{Pereira~dos Santos}},
  \bibinfo{author}{\bibfnamefont{P.}~\bibnamefont{Rosenbusch}},
  \bibinfo{author}{\bibfnamefont{P.}~\bibnamefont{Lemonde}},
  \bibinfo{author}{\bibfnamefont{G.}~\bibnamefont{Santarelli}},
  \bibinfo{author}{\bibfnamefont{P.}~\bibnamefont{Wolf}},
  \bibinfo{author}{\bibfnamefont{A.}~\bibnamefont{Clairon}},
  \bibinfo{author}{\bibfnamefont{A.}~\bibnamefont{Luiten}},
  \bibinfo{author}{\bibfnamefont{M.}~\bibnamefont{Tobar}}, \bibnamefont{and}
  \bibinfo{author}{\bibfnamefont{C.}~\bibnamefont{Salomon}},
 \bibinfo{title}{``Cold atom clocks and applications"}, \bibinfo{journal}{J. Phys. B}
  \textbf{\bibinfo{volume}{38}}, \bibinfo{pages}{S449} (\bibinfo{year}{2005}).

\bibitem[{\citenamefont{Hollberg et~al.}(2005)\citenamefont{Hollberg, Oates,
  Wilpers, Hoyt, Barber, Diddams, Oskay, and Bergquist}}]{Hollberg2005ofw}
\bibinfo{author}{\bibfnamefont{L.}~\bibnamefont{Hollberg}},
  \bibinfo{author}{\bibfnamefont{C.}~\bibnamefont{Oates}},
  \bibinfo{author}{\bibfnamefont{G.}~\bibnamefont{Wilpers}},
  \bibinfo{author}{\bibfnamefont{C.}~\bibnamefont{Hoyt}},
  \bibinfo{author}{\bibfnamefont{Z.}~\bibnamefont{Barber}},
  \bibinfo{author}{\bibfnamefont{S.}~\bibnamefont{Diddams}},
  \bibinfo{author}{\bibfnamefont{W.}~\bibnamefont{Oskay}}, \bibnamefont{and}
  \bibinfo{author}{\bibfnamefont{J.}~\bibnamefont{Bergquist}},  \bibinfo{title}{``Optical frequency/wavelength references"},
  \bibinfo{journal}{J. Phys. B} \textbf{\bibinfo{volume}{38}},
  \bibinfo{pages}{S469} (\bibinfo{year}{2005}).

\bibitem[{\citenamefont{Ketterle and van Druten}(1997)}]{Ketterle1997eco}
\bibinfo{author}{\bibfnamefont{W.}~\bibnamefont{Ketterle}} \bibnamefont{and}
  \bibinfo{author}{\bibfnamefont{N.~J.} \bibnamefont{van Druten}}, \bibinfo{title}{``Evaporative cooling of trapped atoms"},
  \bibinfo{journal}{Adv. At. Mol. Opt. Phys.} \textbf{\bibinfo{volume}{96}},
  \bibinfo{pages}{181} (\bibinfo{year}{1997}).

\bibitem[{\citenamefont{Anderson et~al.}(1995)\citenamefont{Anderson, Ensher,
  Matthews, Wieman, and Cornell}}]{Anderson1995oob}
\bibinfo{author}{\bibfnamefont{M.~H.} \bibnamefont{Anderson}},
  \bibinfo{author}{\bibfnamefont{J.~R.} \bibnamefont{Ensher}},
  \bibinfo{author}{\bibfnamefont{M.~R.} \bibnamefont{Matthews}},
  \bibinfo{author}{\bibfnamefont{C.~E.} \bibnamefont{Wieman}},
  \bibnamefont{and} \bibinfo{author}{\bibfnamefont{E.~A.}
  \bibnamefont{Cornell}}, \bibinfo{title}{``Observation of Bose-Einstein condensation in dilute atomic vapor"}, \bibinfo{journal}{Science}
  \textbf{\bibinfo{volume}{269}}, \bibinfo{pages}{198} (\bibinfo{year}{1995}).

\bibitem[{\citenamefont{Bradley et~al.}(1995)\citenamefont{Bradley, Sackett,
  Tollett, and Hulet}}]{Bradley1995eob}
\bibinfo{author}{\bibfnamefont{C.~C.} \bibnamefont{Bradley}},
  \bibinfo{author}{\bibfnamefont{C.~A.} \bibnamefont{Sackett}},
  \bibinfo{author}{\bibfnamefont{J.~J.} \bibnamefont{Tollett}},
  \bibnamefont{and} \bibinfo{author}{\bibfnamefont{R.~G.} \bibnamefont{Hulet}}, \bibinfo{title}{``Evidence of Bose-Einstein condensation in an atomic gas with attractive interactions"},
  \bibinfo{journal}{Phys. Rev. Lett.} \textbf{\bibinfo{volume}{75}},
  \bibinfo{pages}{1687} (\bibinfo{year}{1995}).

\bibitem[{\citenamefont{Davis et~al.}(1995)\citenamefont{Davis, Mewes, Andrews,
  van Druten, Durfee, Kurn, and Ketterle}}]{Davis1995bec}
\bibinfo{author}{\bibfnamefont{K.~B.} \bibnamefont{Davis}},
  \bibinfo{author}{\bibfnamefont{M.~O.} \bibnamefont{Mewes}},
  \bibinfo{author}{\bibfnamefont{M.~R.} \bibnamefont{Andrews}},
  \bibinfo{author}{\bibfnamefont{N.~J.} \bibnamefont{van Druten}},
  \bibinfo{author}{\bibfnamefont{D.~S.} \bibnamefont{Durfee}},
  \bibinfo{author}{\bibfnamefont{D.~M.} \bibnamefont{Kurn}}, \bibnamefont{and}
  \bibinfo{author}{\bibfnamefont{W.}~\bibnamefont{Ketterle}}, \bibinfo{title}{``Bose-Einstein condensation in a gas of sodium atoms"},
  \bibinfo{journal}{Phys. Rev. Lett.} \textbf{\bibinfo{volume}{75}},
  \bibinfo{pages}{3969} (\bibinfo{year}{1995}).

\bibitem[{\citenamefont{DeMarco and Jin}(1999)}]{DeMarco1999oof}
\bibinfo{author}{\bibfnamefont{B.}~\bibnamefont{DeMarco}} \bibnamefont{and}
  \bibinfo{author}{\bibfnamefont{D.~S.} \bibnamefont{Jin}}, \bibinfo{title}{``Onset of Fermi degeneracy in a trapped atomic gas"},
  \bibinfo{journal}{Science} \textbf{\bibinfo{volume}{285}},
  \bibinfo{pages}{1703} (\bibinfo{year}{1999}).

\bibitem[{\citenamefont{Schreck et~al.}(2001)\citenamefont{Schreck, Khaykovich,
  Corwin, Ferrari, Bourdel, Cubizolles, and Salomon}}]{Schreck2001qbe}
\bibinfo{author}{\bibfnamefont{F.}~\bibnamefont{Schreck}},
  \bibinfo{author}{\bibfnamefont{L.}~\bibnamefont{Khaykovich}},
  \bibinfo{author}{\bibfnamefont{K.~L.} \bibnamefont{Corwin}},
  \bibinfo{author}{\bibfnamefont{G.}~\bibnamefont{Ferrari}},
  \bibinfo{author}{\bibfnamefont{T.}~\bibnamefont{Bourdel}},
  \bibinfo{author}{\bibfnamefont{J.}~\bibnamefont{Cubizolles}},
  \bibnamefont{and} \bibinfo{author}{\bibfnamefont{C.}~\bibnamefont{Salomon}}, \bibinfo{title}{``Quasipure Bose-Einstein condensate immersed in a Fermi sea"},
  \bibinfo{journal}{Phys. Rev. Lett.} \textbf{\bibinfo{volume}{87}},
  \bibinfo{pages}{080403} (\bibinfo{year}{2001}).

\bibitem[{\citenamefont{Truscott et~al.}(2001)\citenamefont{Truscott, Strecker,
  McAlexander, Partridge, and Hulet}}]{Truscott2001oof}
\bibinfo{author}{\bibfnamefont{A.~G.} \bibnamefont{Truscott}},
  \bibinfo{author}{\bibfnamefont{K.~E.} \bibnamefont{Strecker}},
  \bibinfo{author}{\bibfnamefont{W.~I.} \bibnamefont{McAlexander}},
  \bibinfo{author}{\bibfnamefont{G.~B.} \bibnamefont{Partridge}},
  \bibnamefont{and} \bibinfo{author}{\bibfnamefont{R.~G.} \bibnamefont{Hulet}}, \bibinfo{title}{``Observation of Fermi pressure in a gas of trapped atoms"},
  \bibinfo{journal}{Science} \textbf{\bibinfo{volume}{291}},
  \bibinfo{pages}{2570} (\bibinfo{year}{2001}).

\bibitem[{\citenamefont{Inguscio et~al.}(1999)\citenamefont{Inguscio,
  Stringari, and Wieman}}]{Varenna1998}
\bibinfo{editor}{\bibfnamefont{M.}~\bibnamefont{Inguscio}},
  \bibinfo{editor}{\bibfnamefont{S.}~\bibnamefont{Stringari}},
  \bibnamefont{and} \bibinfo{editor}{\bibfnamefont{C.~E.}
  \bibnamefont{Wieman}}, eds., \emph{\bibinfo{title}{Bose-Einstein Condensation
  in Atomic Gases}} (\bibinfo{publisher}{IOS Press, Amsterdam},
  \bibinfo{year}{1999}), \bibinfo{note}{{P}roceedings of the International
  School of Physics ``Enrico Fermi'', Course CXL, Varenna, 7-17 July 1998}.

\bibitem[{\citenamefont{Inguscio et~al.}(2008)\citenamefont{Inguscio, Ketterle,
  and Salomon}}]{Varenna2006}
\bibinfo{editor}{\bibfnamefont{M.}~\bibnamefont{Inguscio}},
  \bibinfo{editor}{\bibfnamefont{W.}~\bibnamefont{Ketterle}}, \bibnamefont{and}
  \bibinfo{editor}{\bibfnamefont{C.}~\bibnamefont{Salomon}}, eds.,
  \emph{\bibinfo{title}{Ultracold Fermi Gases}} (\bibinfo{publisher}{IOS Press,
  Amsterdam}, \bibinfo{year}{2008}), \bibinfo{note}{Proceedings of the
  International School of Physics ``Enrico Fermi'', Course CLXIV, Varenna,
  20-30 June 2006}.

\bibitem[{\citenamefont{Dalfovo et~al.}(1999)\citenamefont{Dalfovo, Giorgini,
  Pitaevskii, and Stringari}}]{Dalfovo1999tob}
\bibinfo{author}{\bibfnamefont{F.}~\bibnamefont{Dalfovo}},
  \bibinfo{author}{\bibfnamefont{S.}~\bibnamefont{Giorgini}},
  \bibinfo{author}{\bibfnamefont{L.~P.} \bibnamefont{Pitaevskii}},
  \bibnamefont{and}
  \bibinfo{author}{\bibfnamefont{S.}~\bibnamefont{Stringari}}, \bibinfo{title}{``Theory of Bose-Einstein condensation in trapped gases"},
  \bibinfo{journal}{Rev. Mod. Phys.} \textbf{\bibinfo{volume}{71}},
  \bibinfo{pages}{463} (\bibinfo{year}{1999}).

\bibitem[{\citenamefont{Giorgini et~al.}(2007)\citenamefont{Giorgini,
  Pitaevskii, and Stringari}}]{Giorgini2007tou}
\bibinfo{author}{\bibfnamefont{S.}~\bibnamefont{Giorgini}},
  \bibinfo{author}{\bibfnamefont{L.~P.} \bibnamefont{Pitaevskii}},
  \bibnamefont{and}
  \bibinfo{author}{\bibfnamefont{S.}~\bibnamefont{Stringari}}, \bibinfo{title}{``Theory of ultracold Fermi gases"},
  \bibinfo{journal}{preprint available at http://arxiv.org/abs/0706.3360}.

\bibitem[{\citenamefont{Stringari and Pitaevskii}(2003)}]{stringaribook}
\bibinfo{author}{\bibfnamefont{S.}~\bibnamefont{Stringari}} \bibnamefont{and}
  \bibinfo{author}{\bibfnamefont{L.}~\bibnamefont{Pitaevskii}},
  \emph{\bibinfo{title}{Bose-Einstein condensation}}
  (\bibinfo{publisher}{Oxford University Press, London}, \bibinfo{year}{2003}).

\bibitem[{\citenamefont{Pethick and Smith}(2008)}]{pethickbook}
\bibinfo{author}{\bibfnamefont{C.~J.} \bibnamefont{Pethick}} \bibnamefont{and}
  \bibinfo{author}{\bibfnamefont{H.}~\bibnamefont{Smith}},
  \emph{\bibinfo{title}{Bose-Einstein condensation in dilute gases}}
  (\bibinfo{publisher}{Cambridge University Press}, \bibinfo{year}{2008}).

\bibitem[{\citenamefont{Grimm et~al.}(2000)\citenamefont{Grimm, Weidem\"uller,
  and Ovchinnikov}}]{Grimm2000odt}
\bibinfo{author}{\bibfnamefont{R.}~\bibnamefont{Grimm}},
  \bibinfo{author}{\bibfnamefont{M.}~\bibnamefont{Weidem\"uller}},
  \bibnamefont{and} \bibinfo{author}{\bibfnamefont{Yu.~B.}
  \bibnamefont{Ovchinnikov}}, \bibinfo{title}{``Optical dipole traps for neutral atoms"}, \bibinfo{journal}{Adv. At. Mol. Opt. Phys.}
  \textbf{\bibinfo{volume}{42}}, \bibinfo{pages}{95} (\bibinfo{year}{2000}).

\bibitem[{\citenamefont{Bloch}(2005)}]{Bloch2005uqg}
\bibinfo{author}{\bibfnamefont{I.}~\bibnamefont{Bloch}}, \bibinfo{title}{``Ultracold quantum gases in optical lattices"},
  \bibinfo{journal}{Nature Phys.} \textbf{\bibinfo{volume}{1}},
  \bibinfo{pages}{23} (\bibinfo{year}{2005}).

\bibitem[{\citenamefont{Greiner and F\"olling}(2008)}]{Greiner2008cmp}
\bibinfo{author}{\bibfnamefont{M.}~\bibnamefont{Greiner}} \bibnamefont{and}
  \bibinfo{author}{\bibfnamefont{S.}~\bibnamefont{F\"olling}}, \bibinfo{title}{``Condensed-matter physics: Optical lattices"},
  \bibinfo{journal}{Nature} \textbf{\bibinfo{volume}{453}},
  \bibinfo{pages}{736} (\bibinfo{year}{2008}).

\bibitem[{\citenamefont{Donley et~al.}(2002)\citenamefont{Donley, Clausen,
  Thompson, and Wieman}}]{Donley2002amc}
\bibinfo{author}{\bibfnamefont{E.~A.} \bibnamefont{Donley}},
  \bibinfo{author}{\bibfnamefont{N.~R.} \bibnamefont{Clausen}},
  \bibinfo{author}{\bibfnamefont{S.~T.} \bibnamefont{Thompson}},
  \bibnamefont{and} \bibinfo{author}{\bibfnamefont{C.~E.}
  \bibnamefont{Wieman}}, \bibinfo{title}{``Atom-molecule coherence in a Bose-Einstein condensate"}, \bibinfo{journal}{Nature}
  \textbf{\bibinfo{volume}{417}}, \bibinfo{pages}{529} (\bibinfo{year}{2002}).

\bibitem[{\citenamefont{Herbig et~al.}(2003)\citenamefont{Herbig, Kraemer,
  Mark, Weber, Chin, N\"agerl, and Grimm}}]{Herbig2003poa}
\bibinfo{author}{\bibfnamefont{J.}~\bibnamefont{Herbig}},
  \bibinfo{author}{\bibfnamefont{T.}~\bibnamefont{Kraemer}},
  \bibinfo{author}{\bibfnamefont{M.}~\bibnamefont{Mark}},
  \bibinfo{author}{\bibfnamefont{T.}~\bibnamefont{Weber}},
  \bibinfo{author}{\bibfnamefont{C.}~\bibnamefont{Chin}},
  \bibinfo{author}{\bibfnamefont{H.-C.} \bibnamefont{N\"agerl}},
  \bibnamefont{and} \bibinfo{author}{\bibfnamefont{R.}~\bibnamefont{Grimm}}, \bibinfo{title}{``Preparation of a pure molecular quantum gas"},
  \bibinfo{journal}{Science} \textbf{\bibinfo{volume}{301}},
  \bibinfo{pages}{1510} (\bibinfo{year}{2003}).

\bibitem[{\citenamefont{D\"urr et~al.}(2004)\citenamefont{D\"urr, Volz, Marte,
  and Rempe}}]{Durr2004oom}
\bibinfo{author}{\bibfnamefont{S.}~\bibnamefont{D\"urr}},
  \bibinfo{author}{\bibfnamefont{T.}~\bibnamefont{Volz}},
  \bibinfo{author}{\bibfnamefont{A.}~\bibnamefont{Marte}}, \bibnamefont{and}
  \bibinfo{author}{\bibfnamefont{G.}~\bibnamefont{Rempe}}, \bibinfo{title}{``Observation of molecules produced from a Bose-Einstein condensate"},
  \bibinfo{journal}{Phys. Rev. Lett.} \textbf{\bibinfo{volume}{92}},
  \bibinfo{eid}{020406} (\bibinfo{year}{2004}).

\bibitem[{\citenamefont{Xu et~al.}(2003)\citenamefont{Xu, Mukaiyama,
  Abo-Shaeer, Chin, Miller, and Ketterle}}]{Xu2003foq}
\bibinfo{author}{\bibfnamefont{K.}~\bibnamefont{Xu}},
  \bibinfo{author}{\bibfnamefont{T.}~\bibnamefont{Mukaiyama}},
  \bibinfo{author}{\bibfnamefont{J.~R.} \bibnamefont{Abo-Shaeer}},
  \bibinfo{author}{\bibfnamefont{J.~K.} \bibnamefont{Chin}},
  \bibinfo{author}{\bibfnamefont{D.~E.} \bibnamefont{Miller}},
  \bibnamefont{and} \bibinfo{author}{\bibfnamefont{W.}~\bibnamefont{Ketterle}}, \bibinfo{title}{``Formation of quantum-degenerate sodium molecules"},
  \bibinfo{journal}{Phys. Rev. Lett.} \textbf{\bibinfo{volume}{91}},
  \bibinfo{eid}{210402} (\bibinfo{year}{2003}).

\bibitem[{\citenamefont{Regal et~al.}(2003)\citenamefont{Regal, Ticknor, Bohn,
  and Jin}}]{Regal2003cum}
\bibinfo{author}{\bibfnamefont{C.~A.} \bibnamefont{Regal}},
  \bibinfo{author}{\bibfnamefont{C.}~\bibnamefont{Ticknor}},
  \bibinfo{author}{\bibfnamefont{J.~L.} \bibnamefont{Bohn}}, \bibnamefont{and}
  \bibinfo{author}{\bibfnamefont{D.~S.} \bibnamefont{Jin}}, \bibinfo{title}{``Creation of ultracold molecules from a Fermi gas of atoms"},
  \bibinfo{journal}{Nature} \textbf{\bibinfo{volume}{424}}, \bibinfo{pages}{47}
  (\bibinfo{year}{2003}).

\bibitem[{\citenamefont{Cubizolles et~al.}(2003)\citenamefont{Cubizolles,
  Bourdel, Kokkelmans, Shlyapnikov, and Salomon}}]{Cubizolles2003pol}
\bibinfo{author}{\bibfnamefont{J.}~\bibnamefont{Cubizolles}},
  \bibinfo{author}{\bibfnamefont{T.}~\bibnamefont{Bourdel}},
  \bibinfo{author}{\bibfnamefont{S.~J. J. M.~F.} \bibnamefont{Kokkelmans}},
  \bibinfo{author}{\bibfnamefont{G.~V.} \bibnamefont{Shlyapnikov}},
  \bibnamefont{and} \bibinfo{author}{\bibfnamefont{C.}~\bibnamefont{Salomon}}, \bibinfo{title}{``Production of long-lived ultracold Li$_2$ molecules from a Fermi gas"},
  \bibinfo{journal}{Phys. Rev. Lett.} \textbf{\bibinfo{volume}{91}},
  \bibinfo{pages}{240401} (\bibinfo{year}{2003}).

\bibitem[{\citenamefont{Jochim et~al.}(2003{\natexlab{a}})\citenamefont{Jochim,
  Bartenstein, Altmeyer, Hendl, Chin, {Hecker Denschlag}, and
  Grimm}}]{Jochim2003pgo}
\bibinfo{author}{\bibfnamefont{S.}~\bibnamefont{Jochim}},
  \bibinfo{author}{\bibfnamefont{M.}~\bibnamefont{Bartenstein}},
  \bibinfo{author}{\bibfnamefont{A.}~\bibnamefont{Altmeyer}},
  \bibinfo{author}{\bibfnamefont{G.}~\bibnamefont{Hendl}},
  \bibinfo{author}{\bibfnamefont{C.}~\bibnamefont{Chin}},
  \bibinfo{author}{\bibfnamefont{J.}~\bibnamefont{{Hecker Denschlag}}},
  \bibnamefont{and} \bibinfo{author}{\bibfnamefont{R.}~\bibnamefont{Grimm}}, \bibinfo{title}{``Pure gas of optically trapped molecules created from fermionic atoms"},
  \bibinfo{journal}{Phys. Rev. Lett} \textbf{\bibinfo{volume}{91}},
  \bibinfo{pages}{240402} (\bibinfo{year}{2003}{\natexlab{a}}).

\bibitem[{\citenamefont{Strecker et~al.}(2003)\citenamefont{Strecker,
  Partridge, and Hulet}}]{Strecker2003coa}
\bibinfo{author}{\bibfnamefont{K.~E.} \bibnamefont{Strecker}},
  \bibinfo{author}{\bibfnamefont{G.~B.} \bibnamefont{Partridge}},
  \bibnamefont{and} \bibinfo{author}{\bibfnamefont{R.~G.} \bibnamefont{Hulet}}, \bibinfo{title}{``Conversion of an atomic Fermi gas to a long-lived molecular Bose gas"},
  \bibinfo{journal}{Phys. Rev. Lett.} \textbf{\bibinfo{volume}{91}},
  \bibinfo{pages}{080406} (\bibinfo{year}{2003}).

\bibitem[{\citenamefont{Wynar et~al.}(2000)\citenamefont{Wynar, {F}reeland,
  {H}an, {R}yu, and {H}einzen}}]{Wynar2000mia}
\bibinfo{author}{\bibfnamefont{R.}~\bibnamefont{Wynar}},
  \bibinfo{author}{\bibfnamefont{R.~S.} \bibnamefont{{F}reeland}},
  \bibinfo{author}{\bibfnamefont{D.~J.} \bibnamefont{{H}an}},
  \bibinfo{author}{\bibfnamefont{C.}~\bibnamefont{{R}yu}}, \bibnamefont{and}
  \bibinfo{author}{\bibfnamefont{D.~J.} \bibnamefont{{H}einzen}}, \bibinfo{title}{``Molecules in a Bose-Einstein condensate"},
  \bibinfo{journal}{Science} \textbf{\bibinfo{volume}{287}},
  \bibinfo{pages}{1016} (\bibinfo{year}{2000}).

\bibitem[{\citenamefont{Greiner et~al.}(2003)\citenamefont{Greiner, Regal, and
  Jin}}]{Greiner2003eoa}
\bibinfo{author}{\bibfnamefont{M.}~\bibnamefont{Greiner}},
  \bibinfo{author}{\bibfnamefont{C.~A.} \bibnamefont{Regal}}, \bibnamefont{and}
  \bibinfo{author}{\bibfnamefont{D.~S.} \bibnamefont{Jin}}, \bibinfo{title}{``Emergence of a molecular Bose-Einstein condensate from a Fermi gas"},
  \bibinfo{journal}{Nature} \textbf{\bibinfo{volume}{426}},
  \bibinfo{pages}{537} (\bibinfo{year}{2003}).

\bibitem[{\citenamefont{Jochim et~al.}(2003)\citenamefont{Jochim,
  Bartenstein, Altmeyer, Hendl, Riedl, Chin, {Hecker Denschlag}, and
  Grimm}}]{Jochim2003bec}
\bibinfo{author}{\bibfnamefont{S.}~\bibnamefont{Jochim}},
  \bibinfo{author}{\bibfnamefont{M.}~\bibnamefont{Bartenstein}},
  \bibinfo{author}{\bibfnamefont{A.}~\bibnamefont{Altmeyer}},
  \bibinfo{author}{\bibfnamefont{G.}~\bibnamefont{Hendl}},
  \bibinfo{author}{\bibfnamefont{S.}~\bibnamefont{Riedl}},
  \bibinfo{author}{\bibfnamefont{C.}~\bibnamefont{Chin}},
  \bibinfo{author}{\bibfnamefont{J.}~\bibnamefont{{Hecker Denschlag}}},
  \bibnamefont{and} \bibinfo{author}{\bibfnamefont{R.}~\bibnamefont{Grimm}}, \bibinfo{title}{``Bose-Einstein condensation of molecules"},
  \bibinfo{journal}{Science} \textbf{\bibinfo{volume}{302}},
  \bibinfo{pages}{2101} (\bibinfo{year}{2003}).

\bibitem[{\citenamefont{Zwierlein et~al.}(2003)\citenamefont{Zwierlein, Stan,
  Schunck, Raupach, Gupta, Hadzibabic, and Ketterle}}]{Zwierlein2003oob}
\bibinfo{author}{\bibfnamefont{M.~W.} \bibnamefont{Zwierlein}},
  \bibinfo{author}{\bibfnamefont{C.~A.} \bibnamefont{Stan}},
  \bibinfo{author}{\bibfnamefont{C.~H.} \bibnamefont{Schunck}},
  \bibinfo{author}{\bibfnamefont{S.~M.~F.} \bibnamefont{Raupach}},
  \bibinfo{author}{\bibfnamefont{S.}~\bibnamefont{Gupta}},
  \bibinfo{author}{\bibfnamefont{Z.}~\bibnamefont{Hadzibabic}},
  \bibnamefont{and} \bibinfo{author}{\bibfnamefont{W.}~\bibnamefont{Ketterle}}, \bibinfo{title}{``Observation of Bose-Einstein condensation of molecules"},
  \bibinfo{journal}{Phys. Rev. Lett.} \textbf{\bibinfo{volume}{91}},
  \bibinfo{pages}{250401} (\bibinfo{year}{2003}).

\bibitem[{\citenamefont{Ospelkaus et~al.}(2006)\citenamefont{Ospelkaus,
  Ospelkaus, Humbert, Ernst, Sengstock, and Bongs}}]{Ospelkaus2006uhm}
\bibinfo{author}{\bibfnamefont{C.}~\bibnamefont{Ospelkaus}},
  \bibinfo{author}{\bibfnamefont{S.}~\bibnamefont{Ospelkaus}},
  \bibinfo{author}{\bibfnamefont{L.}~\bibnamefont{Humbert}},
  \bibinfo{author}{\bibfnamefont{P.}~\bibnamefont{Ernst}},
  \bibinfo{author}{\bibfnamefont{K.}~\bibnamefont{Sengstock}},
  \bibnamefont{and} \bibinfo{author}{\bibfnamefont{K.}~\bibnamefont{Bongs}}, \bibinfo{title}{``Ultracold heteronuclear molecules in a 3D optical lattice"},
  \bibinfo{journal}{Phys. Rev. Lett.} \textbf{\bibinfo{volume}{97}},
  \bibinfo{eid}{120402} (\bibinfo{year}{2006}).

\bibitem[{\citenamefont{Papp and Wieman}(2006)}]{Papp2006ooh}
\bibinfo{author}{\bibfnamefont{S.~B.} \bibnamefont{Papp}} \bibnamefont{and}
  \bibinfo{author}{\bibfnamefont{C.~E.} \bibnamefont{Wieman}}, \bibinfo{title}{``Observation of heteronuclear Feshbach molecules from a $^{85}$Rb-$^{87}$Rb gas"},
  \bibinfo{journal}{Phys. Rev. Lett.} \textbf{\bibinfo{volume}{97}},
  \bibinfo{pages}{180404} (\bibinfo{year}{2006}).

\bibitem[{\citenamefont{Weber et~al.}(2008)\citenamefont{Weber, Barontini,
  Catani, Thalhammer, Inguscio, and Minardi}}]{Weber2008aou}
\bibinfo{author}{\bibfnamefont{C.}~\bibnamefont{Weber}},
  \bibinfo{author}{\bibfnamefont{G.}~\bibnamefont{Barontini}},
  \bibinfo{author}{\bibfnamefont{J.}~\bibnamefont{Catani}},
  \bibinfo{author}{\bibfnamefont{G.}~\bibnamefont{Thalhammer}},
  \bibinfo{author}{\bibfnamefont{M.}~\bibnamefont{Inguscio}}, \bibnamefont{and}
  \bibinfo{author}{\bibfnamefont{F.}~\bibnamefont{Minardi}}, \bibinfo{title}{``Association of ultracold double-species bosonic molecules"},
  \bibinfo{journal}{preprint available at http://arxiv.org/abs/0808.4077}.

\bibitem[{\citenamefont{K\"ohler et~al.}(2006)\citenamefont{K\"ohler, G\'oral,
  and Julienne}}]{Kohler2006poc}
\bibinfo{author}{\bibfnamefont{T.}~\bibnamefont{K\"ohler}},
  \bibinfo{author}{\bibfnamefont{K.}~\bibnamefont{G\'oral}}, \bibnamefont{and}
  \bibinfo{author}{\bibfnamefont{P.~S.} \bibnamefont{Julienne}}, \bibinfo{title}{``Production of cold molecules via magnetically tunable Feshbach resonances"},
  \bibinfo{journal}{Rev. Mod. Phys.} \textbf{\bibinfo{volume}{78}},
  \bibinfo{eid}{1311} (\bibinfo{year}{2006}).

\bibitem[{\citenamefont{Chin et~al.}(2009)\citenamefont{Chin, Grimm, Julienne,
  and Tiesinga}}]{Chin2008fri}
\bibinfo{author}{\bibfnamefont{C.}~\bibnamefont{Chin}},
  \bibinfo{author}{\bibfnamefont{R.}~\bibnamefont{Grimm}},
  \bibinfo{author}{\bibfnamefont{P.}~\bibnamefont{Julienne}}, \bibnamefont{and}
  \bibinfo{author}{\bibfnamefont{E.}~\bibnamefont{Tiesinga}}, \bibinfo{journal}{Rev. Mod. Phys.}, \bibinfo{note}{in preparation}.

\bibitem[{not()}]{notepw}
\bibinfo{note}{Partial waves associated with even angular momentum quantum
  numbers $\ell=0,2,4,6,8$ are by convention denoted as $s$, $d$, $g$, $i$ and $l$-waves,
  respectively. In the case of odd angular momentum quantum numbers
  $\ell=1,3,5,7$ the partial waves are called $p$, $f$, $h$ and $k$-waves,
  respectively. H.~N.~Russell, A.~G.~Shenstone, and L.~A.~Turner, ``Report on notation for atomic spectra", Phys. Rev. \textbf{33}, 900 (1929).}

\bibitem[{\citenamefont{Petrov et~al.}(2004)\citenamefont{Petrov, Salomon, and
  Shlyapnikov}}]{Petrov2004wbm}
\bibinfo{author}{\bibfnamefont{D.~S.} \bibnamefont{Petrov}},
  \bibinfo{author}{\bibfnamefont{C.}~\bibnamefont{Salomon}}, \bibnamefont{and}
  \bibinfo{author}{\bibfnamefont{G.~V.} \bibnamefont{Shlyapnikov}}, \bibinfo{title}{``Weakly bound molecules of fermionic atoms"},
  \bibinfo{journal}{Phys. Rev. Lett.} \textbf{\bibinfo{volume}{93}},
  \bibinfo{pages}{090404} (\bibinfo{year}{2004}).

\bibitem[{\citenamefont{van Abeelen and Verhaar}(1999)}]{vanAbeelen1999tdf}
\bibinfo{author}{\bibfnamefont{F.~A.} \bibnamefont{van Abeelen}}
  \bibnamefont{and} \bibinfo{author}{\bibfnamefont{B.~J.}
  \bibnamefont{Verhaar}}, \bibinfo{title}{``Time-dependent Feshbach resonance scattering and anomalous decay of a Na Bose-Einstein condensate"}, \bibinfo{journal}{Phys. Rev. Lett.}
  \textbf{\bibinfo{volume}{83}}, \bibinfo{pages}{1550} (\bibinfo{year}{1999}).

\bibitem[{\citenamefont{Mies et~al.}(2000)\citenamefont{Mies, Tiesinga, and
  {J}ulienne}}]{Mies2000mof}
\bibinfo{author}{\bibfnamefont{F.~H.} \bibnamefont{Mies}},
  \bibinfo{author}{\bibfnamefont{E.}~\bibnamefont{Tiesinga}},
  \bibnamefont{and} \bibinfo{author}{\bibfnamefont{P.~S.}
  \bibnamefont{{J}ulienne}}, \bibinfo{title}{``Manipulation of Feshbach resonances in ultracold atomic collisions using time-dependent magnetic fields"}, \bibinfo{journal}{Phys. {R}ev. {A}}
  \textbf{\bibinfo{volume}{61}}, \bibinfo{pages}{022721}
  (\bibinfo{year}{2000}).

\bibitem[{\citenamefont{Timmermans et~al.}(1999)\citenamefont{Timmermans,
  Tommasini, Hussein, and Kerman}}]{Timmermans1999fri}
\bibinfo{author}{\bibfnamefont{E.}~\bibnamefont{Timmermans}},
  \bibinfo{author}{\bibfnamefont{P.}~\bibnamefont{Tommasini}},
  \bibinfo{author}{\bibfnamefont{M.}~\bibnamefont{Hussein}}, \bibnamefont{and}
  \bibinfo{author}{\bibfnamefont{A.}~\bibnamefont{Kerman}}, \bibinfo{title}{``Feshbach resonances in atomic Bose-Einstein condensates"},
  \bibinfo{journal}{Phys. Rep.} \textbf{\bibinfo{volume}{315}},
  \bibinfo{pages}{199} (\bibinfo{year}{1999}).

\bibitem[{\citenamefont{Hodby et~al.}(2005)\citenamefont{Hodby, Thompson,
  Regal, Greiner, Wilson, Jin, and Cornell}}]{Hodby2005peo}
\bibinfo{author}{\bibfnamefont{E.}~\bibnamefont{Hodby}},
  \bibinfo{author}{\bibfnamefont{S.~T.} \bibnamefont{Thompson}},
  \bibinfo{author}{\bibfnamefont{C.~A.} \bibnamefont{Regal}},
  \bibinfo{author}{\bibfnamefont{M.}~\bibnamefont{Greiner}},
  \bibinfo{author}{\bibfnamefont{A.~C.} \bibnamefont{Wilson}},
  \bibinfo{author}{\bibfnamefont{D.~S.} \bibnamefont{Jin}}, \bibnamefont{and}
  \bibinfo{author}{\bibfnamefont{E.~A.} \bibnamefont{Cornell}}, \bibinfo{title}{``Production efficiency of ultra-cold Feshbach molecules in bosonic and fermionic systems"},
  \bibinfo{journal}{Phys. Rev. Lett.} \textbf{\bibinfo{volume}{94}},
  \bibinfo{pages}{120402} (\bibinfo{year}{2005}).

\bibitem[{\citenamefont{Hanna et~al.}(2007)\citenamefont{Hanna, K\"{o}hler, and
  Burnett}}]{Hanna2007aom}
\bibinfo{author}{\bibfnamefont{T.~M.} \bibnamefont{Hanna}},
  \bibinfo{author}{\bibfnamefont{T.}~\bibnamefont{K\"{o}hler}},
  \bibnamefont{and} \bibinfo{author}{\bibfnamefont{K.}~\bibnamefont{Burnett}}, \bibinfo{title}{``Association of molecules using a resonantly modulated magnetic field"},
  \bibinfo{journal}{Phys. Rev. A} \textbf{\bibinfo{volume}{75}},
  \bibinfo{eid}{013606} (\bibinfo{year}{2007}).

\bibitem[{\citenamefont{Thompson
  et~al.}(2005{\natexlab{a}})\citenamefont{Thompson, {H}odby, and
  {W}ieman}}]{Thompson2005ump}
\bibinfo{author}{\bibfnamefont{S.~T.} \bibnamefont{Thompson}},
  \bibinfo{author}{\bibfnamefont{E.}~\bibnamefont{Hodby}}, \bibnamefont{and}
  \bibinfo{author}{\bibfnamefont{C.~E.} \bibnamefont{Wieman}}, \bibinfo{title}{``Ultracold molecule production via a resonant oscillating magnetic
    field"},
  \bibinfo{journal}{Phys. {R}ev. {L}ett.} \textbf{\bibinfo{volume}{95}},
  \bibinfo{pages}{190404} (\bibinfo{year}{2005}{\natexlab{a}}).

\bibitem[{\citenamefont{Gaebler et~al.}(2007)\citenamefont{Gaebler, Stewart,
  Bohn, and Jin}}]{Gaebler2007pwf}
\bibinfo{author}{\bibfnamefont{J.~P.} \bibnamefont{Gaebler}},
  \bibinfo{author}{\bibfnamefont{J.~T.} \bibnamefont{Stewart}},
  \bibinfo{author}{\bibfnamefont{J.~L.} \bibnamefont{Bohn}}, \bibnamefont{and}
  \bibinfo{author}{\bibfnamefont{D.~S.} \bibnamefont{Jin}}, \bibinfo{title}{``$p$-wave Feshbach molecules"},
  \bibinfo{journal}{Phys.\ Rev.\ Lett.} \textbf{\bibinfo{volume}{98}},
  \bibinfo{pages}{200403} (\bibinfo{year}{2007}).

\bibitem[{\citenamefont{Zirbel et~al.}(2008{\natexlab{a}})\citenamefont{Zirbel,
  Ni, Ospelkaus, D'Incao, Wieman, Ye, and Jin}}]{Zirbel2008cso}
\bibinfo{author}{\bibfnamefont{J.~J.} \bibnamefont{Zirbel}},
  \bibinfo{author}{\bibfnamefont{K.-K.} \bibnamefont{Ni}},
  \bibinfo{author}{\bibfnamefont{S.}~\bibnamefont{Ospelkaus}},
  \bibinfo{author}{\bibfnamefont{J.~P.} \bibnamefont{D'Incao}},
  \bibinfo{author}{\bibfnamefont{C.~E.} \bibnamefont{Wieman}},
  \bibinfo{author}{\bibfnamefont{J.}~\bibnamefont{Ye}}, \bibnamefont{and}
  \bibinfo{author}{\bibfnamefont{D.~S.} \bibnamefont{Jin}}, \bibinfo{title}{``Collisional stability of fermionic Feshbach molecules"},
  \bibinfo{journal}{Phys.\ Rev.\ Lett.} \textbf{\bibinfo{volume}{100}},
  \bibinfo{pages}{143201} (\bibinfo{year}{2008}{\natexlab{a}}).

\bibitem[{\citenamefont{Thalhammer et~al.}(2006)\citenamefont{Thalhammer,
  Winkler, Lang, Schmid, Grimm, and {Hecker Denschlag}}}]{Thalhammer2006llf}
\bibinfo{author}{\bibfnamefont{G.}~\bibnamefont{Thalhammer}},
  \bibinfo{author}{\bibfnamefont{K.}~\bibnamefont{Winkler}},
  \bibinfo{author}{\bibfnamefont{F.}~\bibnamefont{Lang}},
  \bibinfo{author}{\bibfnamefont{S.}~\bibnamefont{Schmid}},
  \bibinfo{author}{\bibfnamefont{R.}~\bibnamefont{Grimm}}, \bibnamefont{and}
  \bibinfo{author}{\bibfnamefont{J.}~\bibnamefont{{Hecker Denschlag}}}, \bibinfo{title}{``Long-lived Feshbach molecules in a three-dimensional optical lattice"},
  \bibinfo{journal}{Phys. Rev. Lett.} \textbf{\bibinfo{volume}{96}},
  \bibinfo{eid}{050402} (\bibinfo{year}{2006}).

\bibitem[{\citenamefont{Volz et~al.}(2006)\citenamefont{Volz, Syassen, Bauer,
  Hansis, D{\"u}rr, and Rempe}}]{Volz2006pqs}
\bibinfo{author}{\bibfnamefont{T.}~\bibnamefont{Volz}},
  \bibinfo{author}{\bibfnamefont{N.}~\bibnamefont{Syassen}},
  \bibinfo{author}{\bibfnamefont{D.}~\bibnamefont{Bauer}},
  \bibinfo{author}{\bibfnamefont{E.}~\bibnamefont{Hansis}},
  \bibinfo{author}{\bibfnamefont{S.}~\bibnamefont{D{\"u}rr}}, \bibnamefont{and}
  \bibinfo{author}{\bibfnamefont{G.}~\bibnamefont{Rempe}}, \bibinfo{title}{``Preparation of a quantum state with one molecule at each site of an optical lattice"},
  \bibinfo{journal}{Nature Phys.} \textbf{\bibinfo{volume}{2}},
  \bibinfo{pages}{692} (\bibinfo{year}{2006}).

\bibitem[{\citenamefont{Chin and Grimm}(2004)}]{Chin2004tea}
\bibinfo{author}{\bibfnamefont{C.}~\bibnamefont{Chin}} \bibnamefont{and}
  \bibinfo{author}{\bibfnamefont{R.}~\bibnamefont{Grimm}}, \bibinfo{title}{``Thermal equilibrium and efficient evaporation of an ultracold atom-molecule mixture"},
  \bibinfo{journal}{Phys. Rev. A} \textbf{\bibinfo{volume}{69}},
  \bibinfo{eid}{033612} (\bibinfo{year}{2004}).

\bibitem[{\citenamefont{Kokkelmans et~al.}(2004)\citenamefont{Kokkelmans,
  Shlyapnikov, and Salomon}}]{Kokkelmans2004dam}
\bibinfo{author}{\bibfnamefont{S.~J. J. M.~F.} \bibnamefont{Kokkelmans}},
  \bibinfo{author}{\bibfnamefont{G.~V.} \bibnamefont{Shlyapnikov}},
  \bibnamefont{and} \bibinfo{author}{\bibfnamefont{C.}~\bibnamefont{Salomon}}, \bibinfo{title}{``Degenerate atom-molecule mixture in a cold Fermi gas"},
  \bibinfo{journal}{Phys. Rev. A} \textbf{\bibinfo{volume}{69}},
  \bibinfo{eid}{031602} (\bibinfo{year}{2004}).

\bibitem[{\citenamefont{Mark et~al.}(2007{\natexlab{a}})\citenamefont{Mark,
  Ferlaino, Knoop, Danzl, Kraemer, Chin, N\"{a}gerl, and Grimm}}]{Mark2007sou}
\bibinfo{author}{\bibfnamefont{M.}~\bibnamefont{Mark}},
  \bibinfo{author}{\bibfnamefont{F.}~\bibnamefont{Ferlaino}},
  \bibinfo{author}{\bibfnamefont{S.}~\bibnamefont{Knoop}},
  \bibinfo{author}{\bibfnamefont{J.~G.} \bibnamefont{Danzl}},
  \bibinfo{author}{\bibfnamefont{T.}~\bibnamefont{Kraemer}},
  \bibinfo{author}{\bibfnamefont{C.}~\bibnamefont{Chin}},
  \bibinfo{author}{\bibfnamefont{H.-C.} \bibnamefont{N\"{a}gerl}},
  \bibnamefont{and} \bibinfo{author}{\bibfnamefont{R.}~\bibnamefont{Grimm}}, \bibinfo{title}{``Spectroscopy of ultracold trapped cesium Feshbach molecules"},
  \bibinfo{journal}{Phys.\ Rev.\ A} \textbf{\bibinfo{volume}{76}},
  \bibinfo{pages}{042514} (\bibinfo{year}{2007}{\natexlab{a}}).

\bibitem[{\citenamefont{Volz et~al.}(2005)\citenamefont{Volz, D{\"u}rr,
  Syassen, Rempe, van Kempen, and Kokkelmans}}]{Volz2005fso}
\bibinfo{author}{\bibfnamefont{T.}~\bibnamefont{Volz}},
  \bibinfo{author}{\bibfnamefont{S.}~\bibnamefont{D{\"u}rr}},
  \bibinfo{author}{\bibfnamefont{N.}~\bibnamefont{Syassen}},
  \bibinfo{author}{\bibfnamefont{G.}~\bibnamefont{Rempe}},
  \bibinfo{author}{\bibfnamefont{E.}~\bibnamefont{van Kempen}},
  \bibnamefont{and}
  \bibinfo{author}{\bibfnamefont{S.}~\bibnamefont{Kokkelmans}}, \bibinfo{title}{``Feshbach spectroscopy of a shape resonance"},
  \bibinfo{journal}{Phys. Rev. A} \textbf{\bibinfo{volume}{72}},
  \bibinfo{pages}{010704(R)} (\bibinfo{year}{2005}).

\bibitem[{\citenamefont{Mukaiyama et~al.}(2004)\citenamefont{Mukaiyama,
  Abo-Shaeer, Xu, Chin, and Ketterle}}]{Mukaiyama2004dad}
\bibinfo{author}{\bibfnamefont{T.}~\bibnamefont{Mukaiyama}},
  \bibinfo{author}{\bibfnamefont{J.~R.} \bibnamefont{Abo-Shaeer}},
  \bibinfo{author}{\bibfnamefont{K.}~\bibnamefont{Xu}},
  \bibinfo{author}{\bibfnamefont{J.~K.} \bibnamefont{Chin}}, \bibnamefont{and}
  \bibinfo{author}{\bibfnamefont{W.}~\bibnamefont{Ketterle}}, \bibinfo{title}{``Dissociation and decay of ultracold sodium molecules"},
  \bibinfo{journal}{Phys. Rev. Lett.} \textbf{\bibinfo{volume}{92}},
  \bibinfo{pages}{180402} (\bibinfo{year}{2004}).

\bibitem[{\citenamefont{Mark et~al.}(2007)\citenamefont{Mark,
  Kraemer, Waldburger, Herbig, Chin, N\"{a}gerl, and Grimm}}]{Mark2007siw}
\bibinfo{author}{\bibfnamefont{M.}~\bibnamefont{Mark}},
  \bibinfo{author}{\bibfnamefont{T.}~\bibnamefont{Kraemer}},
  \bibinfo{author}{\bibfnamefont{P.}~\bibnamefont{Waldburger}},
  \bibinfo{author}{\bibfnamefont{J.}~\bibnamefont{Herbig}},
  \bibinfo{author}{\bibfnamefont{C.}~\bibnamefont{Chin}},
  \bibinfo{author}{\bibfnamefont{H.-C.} \bibnamefont{N\"{a}gerl}},
  \bibnamefont{and} \bibinfo{author}{\bibfnamefont{R.}~\bibnamefont{Grimm}}, \bibinfo{title}{``St\"uckelberg interferometry with ultracold molecules"},
  \bibinfo{journal}{Phys. Rev. Lett.} \textbf{\bibinfo{volume}{99}},
  \bibinfo{eid}{113201} (\bibinfo{year}{2007}).

\bibitem[{\citenamefont{Knoop et~al.}(2008{\natexlab{a}})\citenamefont{Knoop,
  Mark, Ferlaino, Danzl, Kraemer, N\"{a}gerl, and Grimm}}]{Knoop2008mfm}
\bibinfo{author}{\bibfnamefont{S.}~\bibnamefont{Knoop}},
  \bibinfo{author}{\bibfnamefont{M.}~\bibnamefont{Mark}},
  \bibinfo{author}{\bibfnamefont{F.}~\bibnamefont{Ferlaino}},
  \bibinfo{author}{\bibfnamefont{J.~G.} \bibnamefont{Danzl}},
  \bibinfo{author}{\bibfnamefont{T.}~\bibnamefont{Kraemer}},
  \bibinfo{author}{\bibfnamefont{H.-C.} \bibnamefont{N\"{a}gerl}},
  \bibnamefont{and} \bibinfo{author}{\bibfnamefont{R.}~\bibnamefont{Grimm}}, \bibinfo{title}{``Metastable Feshbach molecules in high rotational states"},
  \bibinfo{journal}{Phys.\ Rev.\ Lett.} \textbf{\bibinfo{volume}{100}},
  \bibinfo{pages}{083002} (\bibinfo{year}{2008}{\natexlab{a}}).

\bibitem[{\citenamefont{Bartenstein et~al.}(2004)\citenamefont{Bartenstein,
  Altmeyer, Riedl, Jochim, Chin, {Hecker Denschlag}, and
  Grimm}}]{Bartenstein2004cfa}
\bibinfo{author}{\bibfnamefont{M.}~\bibnamefont{Bartenstein}},
  \bibinfo{author}{\bibfnamefont{A.}~\bibnamefont{Altmeyer}},
  \bibinfo{author}{\bibfnamefont{S.}~\bibnamefont{Riedl}},
  \bibinfo{author}{\bibfnamefont{S.}~\bibnamefont{Jochim}},
  \bibinfo{author}{\bibfnamefont{C.}~\bibnamefont{Chin}},
  \bibinfo{author}{\bibfnamefont{J.}~\bibnamefont{{Hecker Denschlag}}},
  \bibnamefont{and} \bibinfo{author}{\bibfnamefont{R.}~\bibnamefont{Grimm}}, \bibinfo{title}{``Crossover from a molecular Bose-Einstein condensate to a degenerate Fermi gas"},
  \bibinfo{journal}{Phys. Rev. Lett.} \textbf{\bibinfo{volume}{92}},
  \bibinfo{pages}{120401} (\bibinfo{year}{2004}).

\bibitem[{\citenamefont{Zirbel et~al.}(2008)\citenamefont{Zirbel,
  Ni, Ospelkaus, Nicholson, Olsen, Wieman, Ye, Jin, and
  Julienne}}]{Zirbel2008hmi}
\bibinfo{author}{\bibfnamefont{J.~J.} \bibnamefont{Zirbel}},
  \bibinfo{author}{\bibfnamefont{K.-K.} \bibnamefont{Ni}},
  \bibinfo{author}{\bibfnamefont{S.}~\bibnamefont{Ospelkaus}},
  \bibinfo{author}{\bibfnamefont{T.~L.} \bibnamefont{Nicholson}},
  \bibinfo{author}{\bibfnamefont{M.~L.} \bibnamefont{Olsen}},
  \bibinfo{author}{\bibfnamefont{C.~E.} \bibnamefont{Wieman}},
  \bibinfo{author}{\bibfnamefont{J.}~\bibnamefont{Ye}},
  \bibinfo{author}{\bibfnamefont{D.~S.} \bibnamefont{Jin}}, \bibnamefont{and}
  \bibinfo{author}{\bibfnamefont{P.~S.} \bibnamefont{Julienne}}, \bibinfo{title}{``Heteronuclear molecules in an optical dipole trap"},
  \bibinfo{journal}{Phys. Rev. A} \textbf{\bibinfo{volume}{78}},
  \bibinfo{pages}{013416} (\bibinfo{year}{2008}).

\bibitem[{\citenamefont{Lang et~al.}(2008{\natexlab{a}})\citenamefont{Lang,
  van~der Straten, Brandst\"{a}tter, Thalhammer, Winkler, Julienne, Grimm, and
  Denschlag}}]{Lang2008ctm}
\bibinfo{author}{\bibfnamefont{F.}~\bibnamefont{Lang}},
  \bibinfo{author}{\bibfnamefont{P.}~\bibnamefont{van~der Straten}},
  \bibinfo{author}{\bibfnamefont{B.}~\bibnamefont{Brandst\"{a}tter}},
  \bibinfo{author}{\bibfnamefont{G.}~\bibnamefont{Thalhammer}},
  \bibinfo{author}{\bibfnamefont{K.}~\bibnamefont{Winkler}},
  \bibinfo{author}{\bibfnamefont{P.~S.} \bibnamefont{Julienne}},
  \bibinfo{author}{\bibfnamefont{R.}~\bibnamefont{Grimm}}, \bibnamefont{and}
  \bibinfo{author}{\bibfnamefont{J.} \bibnamefont{{Hecker Denschlag}}}, \bibinfo{title}{``Cruising through molecular bound-state manifolds with radiofrequency"},
  \bibinfo{journal}{Nature Phys.} \textbf{\bibinfo{volume}{4}},
  \bibinfo{pages}{223} (\bibinfo{year}{2008}{\natexlab{a}}).

\bibitem[{\citenamefont{{Le Roy}}(2007)}]{LeRoy2007}
\bibinfo{author}{\bibfnamefont{R.~J.} \bibnamefont{{Le Roy}}},
  \emph{\bibinfo{title}{LEVEL 8.0: A Computer Program for Solving the Radial
  Schrodinger Equation for Bound and Quasibound Levels}}
  (\bibinfo{publisher}{University of Waterloo Chemical Physics Research Report
  CP-663}, \bibinfo{year}{2007}), \bibinfo{note}{see
  http://leroy.uwaterloo.ca/programs/}.

\bibitem[{\citenamefont{Braaten and Hammer}(2006)}]{Braaten2006uif}
\bibinfo{author}{\bibfnamefont{E.}~\bibnamefont{Braaten}} \bibnamefont{and}
  \bibinfo{author}{\bibfnamefont{H.-W.} \bibnamefont{Hammer}}, \bibinfo{title}{``Universality in few-body systems with large scattering length"},
  \bibinfo{journal}{Phys. Rep.} \textbf{\bibinfo{volume}{428}},
  \bibinfo{pages}{259} (\bibinfo{year}{2006}).

\bibitem[{\citenamefont{Jensen et~al.}(2004)\citenamefont{Jensen, Riisager,
  Fedorov, and Garrido}}]{Jensen2004sar}
\bibinfo{author}{\bibfnamefont{A.~S.} \bibnamefont{Jensen}},
  \bibinfo{author}{\bibfnamefont{K.}~\bibnamefont{Riisager}},
  \bibinfo{author}{\bibfnamefont{D.~V.} \bibnamefont{Fedorov}},
  \bibnamefont{and} \bibinfo{author}{\bibfnamefont{E.}~\bibnamefont{Garrido}}, \bibinfo{title}{``Structure and reactions of quantum halos"},
  \bibinfo{journal}{Rev. Mod. Phys.} \textbf{\bibinfo{volume}{76}},
  \bibinfo{pages}{215} (\bibinfo{year}{2004}).

\bibitem[{\citenamefont{Thompson
  et~al.}(2005)\citenamefont{Thompson, Hodby, and
  Wieman}}]{Thompson2005sdo}
\bibinfo{author}{\bibfnamefont{S.~T.} \bibnamefont{Thompson}},
  \bibinfo{author}{\bibfnamefont{E.}~\bibnamefont{Hodby}}, \bibnamefont{and}
  \bibinfo{author}{\bibfnamefont{C.~E.} \bibnamefont{Wieman}}, \bibinfo{title}{``Spontaneous dissociation of $^{85}$Rb Feshbach molecules"},
  \bibinfo{journal}{Phys. Rev. Lett.} \textbf{\bibinfo{volume}{94}},
  \bibinfo{pages}{020401} (\bibinfo{year}{2005}).

\bibitem[{\citenamefont{K\"{o}hler et~al.}(2005)\citenamefont{K\"{o}hler,
  Tiesinga, and Julienne}}]{Kohler2005sdo}
\bibinfo{author}{\bibfnamefont{T.}~\bibnamefont{K\"{o}hler}},
  \bibinfo{author}{\bibfnamefont{E.}~\bibnamefont{Tiesinga}}, \bibnamefont{and}
  \bibinfo{author}{\bibfnamefont{P.~S.} \bibnamefont{Julienne}}, \bibinfo{title}{``Spontaneous dissociation of long-range Feshbach molecules"},
  \bibinfo{journal}{Phys.\ Rev.\ Lett.} \textbf{\bibinfo{volume}{94}},
  \bibinfo{pages}{020402} (\bibinfo{year}{2005}).

\bibitem[{\citenamefont{Braaten and Hammer}(2004)}]{Braaten2004edr}
\bibinfo{author}{\bibfnamefont{E.}~\bibnamefont{Braaten}} \bibnamefont{and}
  \bibinfo{author}{\bibfnamefont{H.-W.} \bibnamefont{Hammer}}, \bibinfo{title}{``Enhanced dimer relaxation in an atomic and molecular Bose-Einstein condensate"},
  \bibinfo{journal}{Phys.\ Rev.\ A} \textbf{\bibinfo{volume}{70}},
  \bibinfo{pages}{042706} (\bibinfo{year}{2004}).

\bibitem[{\citenamefont{D'Incao and Esry}(2005)}]{DIncao2005sls}
\bibinfo{author}{\bibfnamefont{J.~P.} \bibnamefont{D'Incao}} \bibnamefont{and}
  \bibinfo{author}{\bibfnamefont{B.~D.} \bibnamefont{Esry}}, \bibinfo{title}{``Scattering length scaling laws for ultracold three-body collisions"},
  \bibinfo{journal}{Phys.\ Rev.\ Lett.} \textbf{\bibinfo{volume}{94}},
  \bibinfo{pages}{213201} (\bibinfo{year}{2005}).

\bibitem[{\citenamefont{D'Incao and Esry}(2008)}]{DIncao2008som}
\bibinfo{author}{\bibfnamefont{J.~P.} \bibnamefont{D'Incao}} \bibnamefont{and}
  \bibinfo{author}{\bibfnamefont{B.~D.} \bibnamefont{Esry}}, \bibinfo{title}{``Suppression of molecular decay in ultracold gases without Fermi statistics"},
  \bibinfo{journal}{Phys.\ Rev.\ Lett.} \textbf{\bibinfo{volume}{100}},
  \bibinfo{pages}{163201} (\bibinfo{year}{2008}).

\bibitem[{\citenamefont{Ferlaino et~al.}(2008)\citenamefont{Ferlaino, Knoop,
  Mark, Berninger, Sch\"{o}bel, N\"{a}gerl, and Grimm}}]{Ferlaino2008cbt}
\bibinfo{author}{\bibfnamefont{F.}~\bibnamefont{Ferlaino}},
  \bibinfo{author}{\bibfnamefont{S.}~\bibnamefont{Knoop}},
  \bibinfo{author}{\bibfnamefont{M.}~\bibnamefont{Mark}},
  \bibinfo{author}{\bibfnamefont{M.}~\bibnamefont{Berninger}},
  \bibinfo{author}{\bibfnamefont{H.}~\bibnamefont{Sch\"{o}bel}},
  \bibinfo{author}{\bibfnamefont{H.-C.} \bibnamefont{N\"{a}gerl}},
  \bibnamefont{and} \bibinfo{author}{\bibfnamefont{R.}~\bibnamefont{Grimm}}, \bibinfo{title}{``Collisions between tunable halo dimers: exploring an elementary four-body process with identical bosons"},
  \bibinfo{journal}{Phys.\ Rev.\ Lett.} \textbf{\bibinfo{volume}{101}},
  \bibinfo{pages}{023201} (\bibinfo{year}{2008}).

\bibitem[{\citenamefont{Skorniakov and
  Ter-Martirosian}(1957)}]{Skorniakov1957tbp}
\bibinfo{author}{\bibfnamefont{G.~V.} \bibnamefont{Skorniakov}}
  \bibnamefont{and} \bibinfo{author}{\bibfnamefont{K.~A.}
  \bibnamefont{Ter-Martirosian}}, \bibinfo{title}{``Three-body problem for short range forces I. Scattering of low-energy neutrons by deutrons"}, \bibinfo{journal}{Sov.\ Phys.\ JETP}
  \textbf{\bibinfo{volume}{4}}, \bibinfo{pages}{648} (\bibinfo{year}{1957}).

\bibitem[{\citenamefont{Efimov}(1970)}]{Efimov1970ela}
\bibinfo{author}{\bibfnamefont{V.}~\bibnamefont{Efimov}}, \bibinfo{title}{``Energy levels arising from resonant two-body forces in a three-body system"},
  \bibinfo{journal}{Phys. Lett. B} \textbf{\bibinfo{volume}{33}},
  \bibinfo{pages}{563} (\bibinfo{year}{1970}).

\bibitem[{\citenamefont{Efimov}(1971)}]{Efimov1971wbs}
\bibinfo{author}{\bibfnamefont{V.}~\bibnamefont{Efimov}}, \bibinfo{title}{``Weakly-bound states of three resonantly-interacting particles"},
  \bibinfo{journal}{Sov. J. Nucl. Phys.} \textbf{\bibinfo{volume}{12}},
  \bibinfo{pages}{589} (\bibinfo{year}{1971}).

\bibitem[{\citenamefont{Kraemer et~al.}(2006)\citenamefont{Kraemer, Mark,
  Waldburger, Danzl, Chin, Engeser, Lange, Pilch, Jaakkola, N\"agerl
  et~al.}}]{Kraemer2006efe}
\bibinfo{author}{\bibfnamefont{T.}~\bibnamefont{Kraemer}},
  \bibinfo{author}{\bibfnamefont{M.}~\bibnamefont{Mark}},
  \bibinfo{author}{\bibfnamefont{P.}~\bibnamefont{Waldburger}},
  \bibinfo{author}{\bibfnamefont{J.~G.} \bibnamefont{Danzl}},
  \bibinfo{author}{\bibfnamefont{C.}~\bibnamefont{Chin}},
  \bibinfo{author}{\bibfnamefont{B.}~\bibnamefont{Engeser}},
  \bibinfo{author}{\bibfnamefont{A.~D.} \bibnamefont{Lange}},
  \bibinfo{author}{\bibfnamefont{K.}~\bibnamefont{Pilch}},
  \bibinfo{author}{\bibfnamefont{A.}~\bibnamefont{Jaakkola}},
  \bibinfo{author}{\bibfnamefont{H.-C.} \bibnamefont{N\"agerl}}, \bibnamefont{and}
  \bibinfo{author}{\bibfnamefont{R.} \bibnamefont{Grimm}},
\bibinfo{title}{``Evidence for Efimov quantum states in an ultracold gas of caesium atoms"}, \bibinfo{journal}{Nature}
  \textbf{\bibinfo{volume}{440}}, \bibinfo{pages}{315} (\bibinfo{year}{2006}).

\bibitem[{\citenamefont{Knoop et~al.}(2008)\citenamefont{Knoop,
  Ferlaino, Mark, Berninger, Sch\"{o}bel, N\"{a}gerl, and
  Grimm}}]{Knoop2008ooa}
\bibinfo{author}{\bibfnamefont{S.}~\bibnamefont{Knoop}},
  \bibinfo{author}{\bibfnamefont{F.}~\bibnamefont{Ferlaino}},
  \bibinfo{author}{\bibfnamefont{M.}~\bibnamefont{Mark}},
  \bibinfo{author}{\bibfnamefont{M.}~\bibnamefont{Berninger}},
  \bibinfo{author}{\bibfnamefont{H.}~\bibnamefont{Sch\"{o}bel}},
  \bibinfo{author}{\bibfnamefont{H.-C.} \bibnamefont{N\"{a}gerl}},
  \bibnamefont{and} \bibinfo{author}{\bibfnamefont{R.}~\bibnamefont{Grimm}}, \bibinfo{title}{``Observation of an Efimov-like resonance in ultracold atom-dimer scattering"},
  \bibinfo{journal}{preprint available at http://arxiv.org/abs/0807.3306}.

\bibitem[{\citenamefont{Braaten and Hammer}(2001)}]{Braaten2001tbr}
\bibinfo{author}{\bibfnamefont{E.}~\bibnamefont{Braaten}} \bibnamefont{and}
  \bibinfo{author}{\bibfnamefont{H.-W.} \bibnamefont{Hammer}}, \bibinfo{title}{``Three-body recombination into deep bound states in a Bose gas with large scattering length"},
  \bibinfo{journal}{Phys. Rev. Lett.} \textbf{\bibinfo{volume}{87}},
  \bibinfo{pages}{160407} (\bibinfo{year}{2001}).

\bibitem[{\citenamefont{Esry et~al.}(1999)\citenamefont{Esry, Greene, and
  Burke}}]{Esry1999rot}
\bibinfo{author}{\bibfnamefont{B.~D.} \bibnamefont{Esry}},
  \bibinfo{author}{\bibfnamefont{C.~H.} \bibnamefont{Greene}},
  \bibnamefont{and} \bibinfo{author}{\bibfnamefont{J.~P.} \bibnamefont{Burke}}, \bibinfo{title}{``Recombination of three atoms in the ultracold limit"},
  \bibinfo{journal}{Phys. Rev. Lett.} \textbf{\bibinfo{volume}{83}},
  \bibinfo{pages}{1751} (\bibinfo{year}{1999}).

\bibitem[{\citenamefont{Braaten and Hammer}(2007)}]{Braaten2007rdr}
\bibinfo{author}{\bibfnamefont{E.}~\bibnamefont{Braaten}} \bibnamefont{and}
  \bibinfo{author}{\bibfnamefont{H.-W.} \bibnamefont{Hammer}}, \bibinfo{title}{``Resonant dimer relaxation in cold atoms with a large scattering length"},
  \bibinfo{journal}{Phys.\ Rev.\ A} \textbf{\bibinfo{volume}{75}},
  \bibinfo{pages}{052710} (\bibinfo{year}{2007}).

\bibitem[{\citenamefont{Nielsen et~al.}(2002)\citenamefont{Nielsen, Suno, and
  Esry}}]{Nielsen2002eri}
\bibinfo{author}{\bibfnamefont{E.}~\bibnamefont{Nielsen}},
  \bibinfo{author}{\bibfnamefont{H.}~\bibnamefont{Suno}}, \bibnamefont{and}
  \bibinfo{author}{\bibfnamefont{B.~D.} \bibnamefont{Esry}}, \bibinfo{title}{``Efimov resonances in atom-diatom scattering"},
  \bibinfo{journal}{Phys.\ Rev.\ A} \textbf{\bibinfo{volume}{66}},
  \bibinfo{pages}{012705} (\bibinfo{year}{2002}).

\bibitem[{\citenamefont{D'Incao and Esry}(2006)}]{DIncao2006eto}
\bibinfo{author}{\bibfnamefont{J.~P.} \bibnamefont{D'Incao}} \bibnamefont{and}
  \bibinfo{author}{\bibfnamefont{B.~D.} \bibnamefont{Esry}}, \bibinfo{title}{``Enhancing the observability of the Efimov effect in ultracold atomic gas mixtures"},
  \bibinfo{journal}{Phys.\ Rev.\ A} \textbf{\bibinfo{volume}{73}},
  \bibinfo{pages}{030703} (\bibinfo{year}{2006}).

\bibitem[{\citenamefont{Bloch et~al.}(2008)\citenamefont{Bloch, Dalibard, and
  Zwerger}}]{Bloch2008mbp}
\bibinfo{author}{\bibfnamefont{I.}~\bibnamefont{Bloch}},
  \bibinfo{author}{\bibfnamefont{J.}~\bibnamefont{Dalibard}}, \bibnamefont{and}
  \bibinfo{author}{\bibfnamefont{W.}~\bibnamefont{Zwerger}}, \bibinfo{title}{``Many-body physics with ultracold gases"},
  \bibinfo{journal}{Rev.\ Mod.\ Phys.} \textbf{\bibinfo{volume}{80}},
  \bibinfo{pages}{885} (\bibinfo{year}{2008}).

\bibitem[{\citenamefont{Jaksch et~al.}(2002)\citenamefont{Jaksch, Venturi,
  Cirac, Williams, and Zoller}}]{Jaksch2002coa}
\bibinfo{author}{\bibfnamefont{D.}~\bibnamefont{Jaksch}},
  \bibinfo{author}{\bibfnamefont{V.}~\bibnamefont{Venturi}},
  \bibinfo{author}{\bibfnamefont{J.}~\bibnamefont{Cirac}},
  \bibinfo{author}{\bibfnamefont{C.}~\bibnamefont{Williams}}, \bibnamefont{and}
  \bibinfo{author}{\bibfnamefont{P.}~\bibnamefont{Zoller}}, \bibinfo{title}{``Creation of a molecular condensate by dynamically melting a Mott insulator"},
  \bibinfo{journal}{Phys. Rev. Lett.} \textbf{\bibinfo{volume}{89}},
  \bibinfo{pages}{40402} (\bibinfo{year}{2002}).

\bibitem[{\citenamefont{Bergmann et~al.}(1998)\citenamefont{Bergmann, Theuer,
  and Shore}}]{Bergmann1998cpt}
\bibinfo{author}{\bibfnamefont{K.}~\bibnamefont{Bergmann}},
  \bibinfo{author}{\bibfnamefont{H.}~\bibnamefont{Theuer}}, \bibnamefont{and}
  \bibinfo{author}{\bibfnamefont{B.~W.} \bibnamefont{Shore}}, \bibinfo{title}{``Coherent population transfer among quantum states of atoms and molecules"},
  \bibinfo{journal}{Rev. Mod. Phys.} \textbf{\bibinfo{volume}{70}},
  \bibinfo{pages}{1003} (\bibinfo{year}{1998}).

\bibitem[{\citenamefont{Kuklinski et~al.}(1989)\citenamefont{Kuklinski,
  Gaubatz, Hioe, and Bergmann}}]{Kuklinski1989apt}
\bibinfo{author}{\bibfnamefont{J.~R.} \bibnamefont{Kuklinski}},
  \bibinfo{author}{\bibfnamefont{U.}~\bibnamefont{Gaubatz}},
  \bibinfo{author}{\bibfnamefont{F.~T.} \bibnamefont{Hioe}}, \bibnamefont{and}
  \bibinfo{author}{\bibfnamefont{K.}~\bibnamefont{Bergmann}}, \bibinfo{title}{``Adiabatic population transfer in a three-level system driven by delayed laser pulses"},
  \bibinfo{journal}{Phys. Rev. A} \textbf{\bibinfo{volume}{40}},
  \bibinfo{pages}{6741} (\bibinfo{year}{1989}).

\bibitem[{\citenamefont{Oreg et~al.}(1984)\citenamefont{Oreg, Hioe, and
  Eberly}}]{Oreg1984afi}
\bibinfo{author}{\bibfnamefont{J.}~\bibnamefont{Oreg}},
  \bibinfo{author}{\bibfnamefont{F.~T.} \bibnamefont{Hioe}}, \bibnamefont{and}
  \bibinfo{author}{\bibfnamefont{J.~H.} \bibnamefont{Eberly}}, \bibinfo{title}{``Adiabatic following in multilevel systems"},
  \bibinfo{journal}{Phys. Rev. A} \textbf{\bibinfo{volume}{29}},
  \bibinfo{pages}{690} (\bibinfo{year}{1984}).

\bibitem[{\citenamefont{Winkler et~al.}(2007)\citenamefont{Winkler, Lang,
  Thalhammer, v.~d. Straten, Grimm, and {Hecker Denschlag}}}]{Winkler2007cot}
\bibinfo{author}{\bibfnamefont{K.}~\bibnamefont{Winkler}},
  \bibinfo{author}{\bibfnamefont{F.}~\bibnamefont{Lang}},
  \bibinfo{author}{\bibfnamefont{G.}~\bibnamefont{Thalhammer}},
  \bibinfo{author}{\bibfnamefont{P.}~\bibnamefont{v.~d. Straten}},
  \bibinfo{author}{\bibfnamefont{R.}~\bibnamefont{Grimm}}, \bibnamefont{and}
  \bibinfo{author}{\bibfnamefont{J.}~\bibnamefont{{Hecker Denschlag}}}, \bibinfo{title}{``Coherent optical transfer of Feshbach molecules to a lower vibrational state"},
  \bibinfo{journal}{Phys. Rev. Lett.} \textbf{\bibinfo{volume}{98}},
  \bibinfo{eid}{043201} (\bibinfo{year}{2007}).

\bibitem[{\citenamefont{Ospelkaus et~al.}(2008)\citenamefont{Ospelkaus, Pe'er,
  Ni, Zirbel, Neyenhuis, Kotochigova, Julienne, Ye, and
  Jin}}]{Ospelkaus2008est}
\bibinfo{author}{\bibfnamefont{S.}~\bibnamefont{Ospelkaus}},
  \bibinfo{author}{\bibfnamefont{A.}~\bibnamefont{Pe'er}},
  \bibinfo{author}{\bibfnamefont{K.-K.} \bibnamefont{Ni}},
  \bibinfo{author}{\bibfnamefont{J.~J.} \bibnamefont{Zirbel}},
  \bibinfo{author}{\bibfnamefont{B.}~\bibnamefont{Neyenhuis}},
  \bibinfo{author}{\bibfnamefont{S.}~\bibnamefont{Kotochigova}},
  \bibinfo{author}{\bibfnamefont{P.~S.} \bibnamefont{Julienne}},
  \bibinfo{author}{\bibfnamefont{J.}~\bibnamefont{Ye}}, \bibnamefont{and}
  \bibinfo{author}{\bibfnamefont{D.~S.} \bibnamefont{Jin}}, \bibinfo{title}{``Efficient state transfer in an ultracold dense gas of heteronuclear molecules"},
  \bibinfo{journal}{Nature Phys.} \textbf{\bibinfo{volume}{4}},
  \bibinfo{pages}{622} (\bibinfo{year}{2008}).

\bibitem[{\citenamefont{Danzl et~al.}(2008)\citenamefont{Danzl, Haller,
  Gustavsson, Mark, Hart, Bouloufa, Dulieu, Ritsch, and
  N\"agerl}}]{Danzl2008qgo}
\bibinfo{author}{\bibfnamefont{J.~G.} \bibnamefont{Danzl}},
  \bibinfo{author}{\bibfnamefont{E.}~\bibnamefont{Haller}},
  \bibinfo{author}{\bibfnamefont{M.}~\bibnamefont{Gustavsson}},
  \bibinfo{author}{\bibfnamefont{M.~J.} \bibnamefont{Mark}},
  \bibinfo{author}{\bibfnamefont{R.}~\bibnamefont{Hart}},
  \bibinfo{author}{\bibfnamefont{N.}~\bibnamefont{Bouloufa}},
  \bibinfo{author}{\bibfnamefont{O.}~\bibnamefont{Dulieu}},
  \bibinfo{author}{\bibfnamefont{H.}~\bibnamefont{Ritsch}}, \bibnamefont{and}
  \bibinfo{author}{\bibfnamefont{H.-C.} \bibnamefont{N\"agerl}}, \bibinfo{title}{``Quantum gas of deeply bound ground state molecules"},
  \bibinfo{journal}{Science} \textbf{\bibinfo{volume}{321}},
  \bibinfo{pages}{1062} (\bibinfo{year}{2008}).

\bibitem[{\citenamefont{Ni et~al.}(2008)\citenamefont{Ni, Ospelkaus, {de
  Miranda}, Pe'er, Neyenhuis, Zirbel, Kotochigova, Julienne, Jin, and
  Ye}}]{Ni2008ahp}
\bibinfo{author}{\bibfnamefont{K.-K.} \bibnamefont{Ni}},
  \bibinfo{author}{\bibfnamefont{S.}~\bibnamefont{Ospelkaus}},
  \bibinfo{author}{\bibfnamefont{M.~H.~G.} \bibnamefont{{de Miranda}}},
  \bibinfo{author}{\bibfnamefont{A.}~\bibnamefont{Pe'er}},
  \bibinfo{author}{\bibfnamefont{B.}~\bibnamefont{Neyenhuis}},
  \bibinfo{author}{\bibfnamefont{J.~J.} \bibnamefont{Zirbel}},
  \bibinfo{author}{\bibfnamefont{S.}~\bibnamefont{Kotochigova}},
  \bibinfo{author}{\bibfnamefont{P.~S.} \bibnamefont{Julienne}},
  \bibinfo{author}{\bibfnamefont{D.~S.} \bibnamefont{Jin}}, \bibnamefont{and}
  \bibinfo{author}{\bibfnamefont{J.}~\bibnamefont{Ye}},
\bibinfo{title}{``A high phase-space-density gas of polar molecules"},
  \bibinfo{journal}{preprint available at http://arxiv.org/abs/0808.2963}.

\bibitem[{\citenamefont{Lang et~al.}(2008)\citenamefont{Lang,
  Winkler, Strauss, Grimm, and Denschlag}}]{Lang2008umi}
\bibinfo{author}{\bibfnamefont{F.}~\bibnamefont{Lang}},
  \bibinfo{author}{\bibfnamefont{K.}~\bibnamefont{Winkler}},
  \bibinfo{author}{\bibfnamefont{C.}~\bibnamefont{Strauss}},
  \bibinfo{author}{\bibfnamefont{R.}~\bibnamefont{Grimm}}, \bibnamefont{and}
  \bibinfo{author}{\bibfnamefont{J.} \bibnamefont{{Hecker Denschlag}}},
\bibinfo{title}{``Ultracold molecules in the ro-vibrational triplet ground state"},
  \bibinfo{journal}{preprint available at http://arxiv.org/abs/0809.0061}.

\bibitem[{Nae()}]{Naegerl2008}
\bibinfo{note}{H.-C.\ N\"agerl, private communication.}

\bibitem[{\citenamefont{Winkler et~al.}(2006)\citenamefont{Winkler, Thalhammer,
  Lang, Grimm, {Hecker Denschlag}, Daley, Kantian, B{\"u}chler, and
  Zoller}}]{Winkler2006rba}
\bibinfo{author}{\bibfnamefont{K.}~\bibnamefont{Winkler}},
  \bibinfo{author}{\bibfnamefont{G.}~\bibnamefont{Thalhammer}},
  \bibinfo{author}{\bibfnamefont{F.}~\bibnamefont{Lang}},
  \bibinfo{author}{\bibfnamefont{R.}~\bibnamefont{Grimm}},
  \bibinfo{author}{\bibfnamefont{J.}~\bibnamefont{{Hecker Denschlag}}},
  \bibinfo{author}{\bibfnamefont{A.~J.} \bibnamefont{Daley}},
  \bibinfo{author}{\bibfnamefont{A.}~\bibnamefont{Kantian}},
  \bibinfo{author}{\bibfnamefont{H.~P.} \bibnamefont{B{\"u}chler}},
  \bibnamefont{and} \bibinfo{author}{\bibfnamefont{P.}~\bibnamefont{Zoller}}, \bibinfo{title}{``Repulsively bound atom pairs in an optical lattice"},
  \bibinfo{journal}{Nature} \textbf{\bibinfo{volume}{441}},
  \bibinfo{pages}{853} (\bibinfo{year}{2006}).

\bibitem[{\citenamefont{Greiner et~al.}(2002)\citenamefont{Greiner, Mandel,
  Esslinger, H\"ansch, and Bloch}}]{Greiner2002qpt}
\bibinfo{author}{\bibfnamefont{M.}~\bibnamefont{Greiner}},
  \bibinfo{author}{\bibfnamefont{O.}~\bibnamefont{Mandel}},
  \bibinfo{author}{\bibfnamefont{T.}~\bibnamefont{Esslinger}},
  \bibinfo{author}{\bibfnamefont{T.~W.} \bibnamefont{H\"ansch}},
  \bibnamefont{and} \bibinfo{author}{\bibfnamefont{I.}~\bibnamefont{Bloch}}, \bibinfo{title}{``Quantum phase transition from a superfluid to a Mott insulator in a gas of ultracold atoms"},
  \bibinfo{journal}{Nature} \textbf{\bibinfo{volume}{415}}, \bibinfo{pages}{39}
  (\bibinfo{year}{2002}).

\bibitem[{\citenamefont{Moritz et~al.}(2005)\citenamefont{Moritz, St\"oferle,
  G\"unter, K\"ohl, and Esslinger}}]{Moritz2005cim}
\bibinfo{author}{\bibfnamefont{H.}~\bibnamefont{Moritz}},
  \bibinfo{author}{\bibfnamefont{T.}~\bibnamefont{St\"oferle}},
  \bibinfo{author}{\bibfnamefont{K.}~\bibnamefont{G\"unter}},
  \bibinfo{author}{\bibfnamefont{M.}~\bibnamefont{K\"ohl}}, \bibnamefont{and}
  \bibinfo{author}{\bibfnamefont{T.}~\bibnamefont{Esslinger}}, \bibinfo{title}{``Confinement induced molecules in a 1D Fermi gas"},
  \bibinfo{journal}{Phys. Rev. Lett.} \textbf{\bibinfo{volume}{94}},
  \bibinfo{eid}{210401} (\bibinfo{year}{2005}).

\bibitem[{\citenamefont{Syassen et~al.}(2008)\citenamefont{Syassen, Bauer,
  Lettner, Volz, Dietze, Garcia-Ripoll, Cirac, Rempe, and
  Durr}}]{Syassen2008sdi}
\bibinfo{author}{\bibfnamefont{N.}~\bibnamefont{Syassen}},
  \bibinfo{author}{\bibfnamefont{D.~M.} \bibnamefont{Bauer}},
  \bibinfo{author}{\bibfnamefont{M.}~\bibnamefont{Lettner}},
  \bibinfo{author}{\bibfnamefont{T.}~\bibnamefont{Volz}},
  \bibinfo{author}{\bibfnamefont{D.}~\bibnamefont{Dietze}},
  \bibinfo{author}{\bibfnamefont{J.~J.} \bibnamefont{Garcia-Ripoll}},
  \bibinfo{author}{\bibfnamefont{J.~I.} \bibnamefont{Cirac}},
  \bibinfo{author}{\bibfnamefont{G.}~\bibnamefont{Rempe}}, \bibnamefont{and}
  \bibinfo{author}{\bibfnamefont{S.}~\bibnamefont{D\"urr}}, \bibinfo{title}{``Strong dissipation inhibits losses and induces correlations in cold molecular gases"},
  \bibinfo{journal}{Science} \textbf{\bibinfo{volume}{320}},
  \bibinfo{pages}{1329} (\bibinfo{year}{2008}).

\bibitem[{\citenamefont{Taglieber et~al.}(2008)\citenamefont{Taglieber, Voigt,
  Aoki, H\"{a}nsch, and Dieckmann}}]{Taglieber2008qdt}
\bibinfo{author}{\bibfnamefont{M.}~\bibnamefont{Taglieber}},
  \bibinfo{author}{\bibfnamefont{A.-C.} \bibnamefont{Voigt}},
  \bibinfo{author}{\bibfnamefont{T.}~\bibnamefont{Aoki}},
  \bibinfo{author}{\bibfnamefont{T.~W.} \bibnamefont{H\"{a}nsch}},
  \bibnamefont{and}
  \bibinfo{author}{\bibfnamefont{K.}~\bibnamefont{Dieckmann}}, \bibinfo{title}{``Quantum degenerate two-species Fermi-Fermi mixture coexisting with a Bose-Einstein condensate"},
  \bibinfo{journal}{Phys. Rev. Lett.} \textbf{\bibinfo{volume}{100}},
  \bibinfo{eid}{010401} (\bibinfo{year}{2008}).

\bibitem[{\citenamefont{Wille et~al.}(2008)\citenamefont{Wille, Spiegelhalder,
  Kerner, Naik, Trenkwalder, Hendl, Schreck, Grimm, Tiecke, Walraven
  et~al.}}]{Wille2008eau}
\bibinfo{author}{\bibfnamefont{E.}~\bibnamefont{Wille}},
  \bibinfo{author}{\bibfnamefont{F.~M.} \bibnamefont{Spiegelhalder}},
  \bibinfo{author}{\bibfnamefont{G.}~\bibnamefont{Kerner}},
  \bibinfo{author}{\bibfnamefont{D.}~\bibnamefont{Naik}},
  \bibinfo{author}{\bibfnamefont{A.}~\bibnamefont{Trenkwalder}},
  \bibinfo{author}{\bibfnamefont{G.}~\bibnamefont{Hendl}},
  \bibinfo{author}{\bibfnamefont{F.}~\bibnamefont{Schreck}},
  \bibinfo{author}{\bibfnamefont{R.}~\bibnamefont{Grimm}},
  \bibinfo{author}{\bibfnamefont{T.~G.} \bibnamefont{Tiecke}},
  \bibinfo{author}{\bibfnamefont{J.~T.~M.} \bibnamefont{Walraven}},
  \bibinfo{author}{\bibfnamefont{S.~J.~J.~M.~F.} \bibnamefont{Kokkelmans}},
  \bibinfo{author}{\bibfnamefont{E.} \bibnamefont{Tiesinga}}, \bibnamefont{and}
  \bibinfo{author}{\bibfnamefont{P.~S.} \bibnamefont{Julienne}},
\bibinfo{title}{``Exploring an ultracold Fermi-Fermi mixture: Interspecies Feshbach resonances and scattering properties of $^6$Li and $^{40}$K"},
\bibinfo{journal}{Phys. Rev. Lett.}
  \textbf{\bibinfo{volume}{100}}, \bibinfo{pages}{053201}
  (\bibinfo{year}{2008}).

\bibitem[{\citenamefont{Chin et~al.}(2005)\citenamefont{Chin, Kraemer, Mark,
  Herbig, Waldburger, N\"agerl, and Grimm}}]{Chin2005oof}
\bibinfo{author}{\bibfnamefont{C.}~\bibnamefont{Chin}},
  \bibinfo{author}{\bibfnamefont{T.}~\bibnamefont{Kraemer}},
  \bibinfo{author}{\bibfnamefont{M.}~\bibnamefont{Mark}},
  \bibinfo{author}{\bibfnamefont{J.}~\bibnamefont{Herbig}},
  \bibinfo{author}{\bibfnamefont{P.}~\bibnamefont{Waldburger}},
  \bibinfo{author}{\bibfnamefont{H.-C.} \bibnamefont{N\"agerl}},
  \bibnamefont{and} \bibinfo{author}{\bibfnamefont{R.}~\bibnamefont{Grimm}}, \bibinfo{title}{``Observation of Feshbach-like resonances in collisions between ultracold molecules"},
  \bibinfo{journal}{Phys. Rev. Lett.} \textbf{\bibinfo{volume}{94}},
  \bibinfo{pages}{123201} (\bibinfo{year}{2005}).

\end{thebibliography}
\end{document}